\documentclass[manuscript]{acmart}
\usepackage[ruled,vlined]{algorithm2e}
\usepackage{algorithmicx}
\usepackage{algpseudocode}
\usepackage{enumitem}
\usepackage{multirow}
\usepackage{bm}
\usepackage{subfig}

\AtBeginDocument{%
  \providecommand\BibTeX{{%
    \normalfont B\kern-0.5em{\scshape i\kern-0.25em b}\kern-0.8em\TeX}}}

\setcopyright{acmcopyright}
\copyrightyear{2018}
\acmYear{2018}
\acmDOI{10.1145/1122445.1122456}

\acmConference[Woodstock '18]{Woodstock '18: ACM Symposium on Neural
  Gaze Detection}{June 03--05, 2018}{Woodstock, NY}
\acmBooktitle{Woodstock '18: ACM Symposium on Neural Gaze Detection,
  June 03--05, 2018, Woodstock, NY}
\acmPrice{15.00}
\acmISBN{978-1-4503-XXXX-X/18/06}

\begin{document}
\title{Causal Disentanglement with Network Information for Debiased
Recommendations}

\author{Paras Sheth}
\affiliation{%
  \institution{Arizona State University}
  %\streetaddress{P.O. Box 1212}
  \city{Tempe}
  \state{AZ}
  \country{USA}
}
\email{psheth5@asu.edu}
\author{Ruocheng Guo}
\email{ruocheng.guo@cityu.edu.hk}
\affiliation{%
  \institution{City University of Hong Kong}
  %\streetaddress{P.O. Box 1212}
  \country{Hong Kong}
}

\author{Lu Cheng}
\email{lcheng35@asu.edu}
\affiliation{%
  \institution{Arizona State University}
  %\streetaddress{P.O. Box 1212}
  \city{Tempe}
  \state{AZ}
  \country{USA}
  }
%\email{larst@affiliation.org}

\author{Huan Liu}
\email{huanliu@asu.edu}
\affiliation{%
  \institution{Arizona State University}
  %\streetaddress{P.O. Box 1212}
  \city{Tempe}
  \state{AZ}
  \country{USA}
  }

\author{K. Sel\c{c}uk Candan}
\email{candan@asu.edu}
\affiliation{%
  \institution{Arizona State University}
  %\streetaddress{P.O. Box 1212}
  \city{Tempe}
  \state{AZ}
  \country{USA}
  }

% \affiliation{%

% \affiliation{%

% \affiliation{%

% \affiliation{%


\begin{abstract}
  Recommender systems aim to recommend new items to users by learning user and item representations. In practice, these representations are highly entangled as they consist of information about multiple factors, including user's interests, item attributes along with confounding factors such as user conformity, and item popularity. Considering these entangled representations for inferring user preference may lead to biased recommendations  (e.g., when the recommender model recommends popular items even if they do not align with the user's interests).
  Recent research proposes to debias by modeling a recommender system from a causal perspective. The exposure and the ratings are analogous to the treatment and the outcome in the causal inference framework, respectively. The critical challenge in this setting is accounting for the hidden confounders. These confounders are unobserved, making it hard to measure them. On the other hand, since these confounders affect both the exposure and the ratings, it is essential to account for them in generating debiased recommendations. To better approximate hidden confounders, we propose to leverage network information (i.e., user-social and user-item networks), which are shown to influence how users discover and interact with an item. Aside from the user conformity, aspects of confounding such as item popularity present in the network information is also captured in our method with the aid of \textit{causal disentanglement} which unravels the learned representations into independent factors that are responsible for (a) modeling the exposure of an item to the user, (b) predicting the ratings, and (c) controlling the hidden confounders. Experiments on real-world datasets validate the effectiveness of the proposed model for debiasing recommender systems.
\end{abstract}





\maketitle

\section{Introduction}
Recommender systems have become ubiquitous in our daily lives. These systems recommend new items to the users based on their past interactions. However, recent studies have shown that these systems suffer from multiple types of biases, including conformity bias and popularity bias~\cite{wang2020causal,schnabel2016recommendations,chen2020bias}. Due to these biases, the performance of these systems deteriorates, and the recommender systems may not recommend relevant items to the users. As a result, these biases hurt user engagement which may harm the business in the long term~\cite{abdollahpouri2021user,zhu2021popularity}. Developing methods to mitigate these biases and improve recommendation performance is a common approach to address this problem~\cite{borges2021mitigating,wei2021model,zhang2021causal,liang2016causal}.

Recently, a series of work investigates how to pose the problem of recommender systems from a causal perspective to mitigate the various types of bias. In recommender systems, the users are first exposed to a set of items based on the selection mechanisms of an existing recommender algorithm. Then, a user interacts with a subset of these items generating the observed ratings from the set of the exposed items. Under the causal lens, the exposure of an item to the user is considered the treatment variable $T$, such that $T=0$ implies the user was not exposed to an item and $T=1$ implies the user was exposed to a particular item. At the same time, the actual ratings are considered to be the outcomes~\cite{wang2020causal,liang2016causal}. Traditional recommender systems only utilize the observed user-item ratings to infer the user preference and recommend new items. However, this strategy only provides unbiased  inference of user preference over items if users randomly interacted with the items. This is because randomly assigned treatment eliminates the bias induced by the confounders~\cite{schulz1998randomized}. Since users do not interact with items at random, inferring their preference from the observed ratings data becomes challenging. Furthermore, in most cases, there exist some hidden confounders -- the variables that affect both the exposure (treatment) and the ratings (outcomes). For instance, recommender systems may recommend popular items to the user, and at the same time, due to herd mentality~\cite{baddeley2010herding}, users may rate these items similarly. Thus, the item's popularity acts as a confounder in this setting. Also, to make valid causal inference, it is essential to adjust for all confounders~\cite{rosenbaum1983central}. Recent works, including~\cite{wang2020causal} and \cite{liang2016causal}, learn to estimate the true exposure using Poisson Factorization models and use the estimated exposure as a substitute for hidden confounders.
%Similar to this setting~\cite{li2021causal} use user's social relations as a substitute for hidden confounders to produce debiased recommendations. 
%Another line of work~\cite{zheng2021disentangling} considers the implicit recommender setting where the user's interest and conformity are the cause for the observed clicks or views. The authors propose a model to disentangle the user and item representations to generate debiased recommendations. Moreover, Yang et al. ~\cite{yang2021top} aims to improve the personalized rankings of recommender systems by applying Pearl's causal inference framework~\cite{peters2017elements}. The authors considered the user profiles as a common cause for the list of items exposed to a user as well as the list of items a user interacts with.%Another work~\cite{yang2021top} aims to improve the personalized rankings of recommender systems by leveraging Pearl's causal inference framework~\cite{peters2017elements} by considering the user profiles as a common cause for the list of items exposed to a user and the list of item a user interacts with.
\begin{figure}[h]
    \centering
    \includegraphics[width=\textwidth]{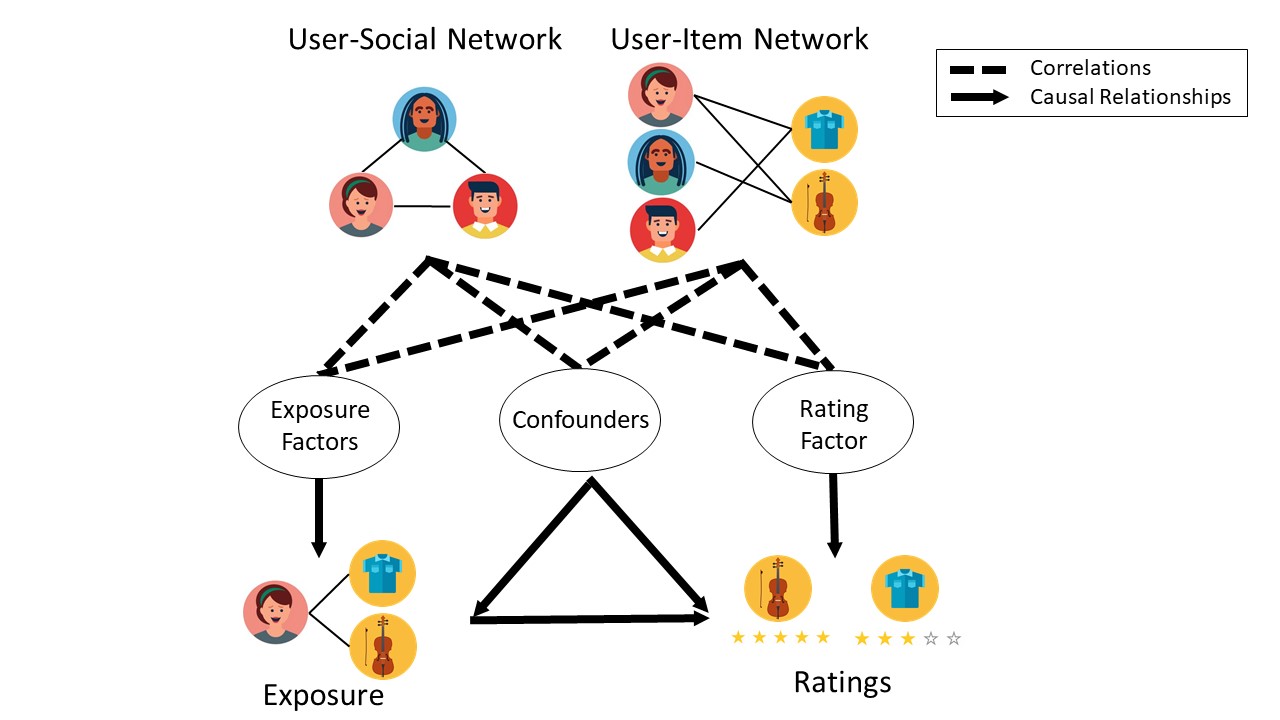}
    \caption{The social recommender systems from a causal perspective. Exposure of an item to a user denotes the treatment; the observed ratings denote the outcomes. The disentangled factors from the user-social networks and item-user networks correlate with the factors for modeling the exposure, factors responsible for the hidden confounders, and factors for predicting the rating.}
    \label{causal_graph}
\end{figure}

 An important problem that has not been investigated is the confounding bias in recommender systems with \textit{social network information}.
 %A crucial feature that has been overlooked by existing literature in causal recommender systems is \textit{social network information}. 
 Social recommender systems learn the user and item representations by leveraging both the user-social networks and user-item networks~\cite{fan2019graph,yang2021consisrec,zafarani2014social}. These representations consist of a variety of information, and not all the information is helpful in terms of mitigating the confounding bias. Under the causal setting, factors such as social relations can act as noisy measures of confounders. For instance, in the physical world a user is likely to seek suggestions from their friends before making any purchases and the user's friends may provide good recommendations~\cite{sinha2001comparing}. A user's preferences are similar to or influenced by their social relations. Analogous to the physical world, this can also be observed in the digital world. This can be corroborated by social correlation theories such as homophily~\cite{ma2009trust} and social influence~\cite{marsden1993network}. Homophily indicates that users with similar preferences are more likely to be connected, and social influence reveals that users who are connected are more likely to have similar preferences. Unfortunately, current social recommendation models cannot distinguish between confounding factors and other factors. Hence, they learn highly entangled representations which lead to biased recommendations. To alleviate this problem, we aim to causally disentangle the learned representations when there exists social network information. This is a challenging task due to the following reasons: 
\begin{itemize}[leftmargin=*]
    \item First, the learned representations consist of a variety of information. Each component of the representation may be effective for either predicting the exposure of an item to the user or acting as a confounder between the exposure and rating, or useful in predicting the rating. For example, user representations learned from complex user interactions might consist of latent features that are highly entangled with each other, such as user preferences and the confounding factor -- user conformity~\cite{zheng2021disentangling}. Similarly, item representations consist of entangled information about an item's latent features and confounding factors, including the item's popularity. Thus, the representations should be carefully disentangled such that each factor has its distinct contribution.
    \item Second, to make valid causal inferences, it is essential that the learned factors are independent of each other~\cite{hassanpour2019learning}. Hence, %while learning the representations, 
    it is critical to ensure independence among the disentangled factors in the learned representations.
    \item Third, the disentangled representations learned for the confounders still exhibit some degree of bias, although less than the entangled representations. This remaining bias exists because, the confounders contribute to predicting both the treatment (exposure) and the outcomes (ratings)~\cite{hassanpour2019counterfactual}. Thus, the remaining bias should also be mitigated. %accounted for.
\end{itemize}
To deal with the above challenges, we propose a causal disentanglement model for social recommender systems\footnote{We will release the code once the paper gets accepted}. %when posing social recommender systems from a causal perspective, we propose a causal disentanglement model for social recommender systems. 
%For example, user representations learned from complex user interactions might consist of latent features that are highly entangled with each other, such as user preferences and the confounding factor -- user conformity~\cite{zheng2021disentangling}. Similarly, item representations consist of entangled information about an item's latent features and confounding factors, including the item's popularity. Since all this information is available in the latent space and at the same time it is highly entangled, we need to disentangle representations into independent factors such that each factor has its unique contributions to mitigating the confounding bias or predicting the ratings. This work explicitly considers two common types of networks -- user-social networks and user-item interaction networks. 
We first formulate the social recommender system problem from a causal perspective as illustrated by the causal graph in Fig. \ref{causal_graph}. To learn the representations, we explicitly consider two common types of networks -- user-social networks and user-item interaction networks. We argue that %instead of learning the entangled representations 
if the representations are disentangled into specific components as highlighted in Fig. \ref{causal_graph}, it would facilitate capturing each component's contribution for a specific task and aid in generating debiased recommendations. %To this end, we propose 
In our causal disentanglement model, the user and item latent representations are learned using the user-social and user-item interaction networks. Then, the representations are disentangled to account for three independent types of variables: confounding variables, factors that model the exposure (treatment) only, and those that are predictive of the ratings (outcome) only. Including the hidden confounders %and considering their contributions while 
in computing the rating predictions would allow us to address the confounding bias. As a result, it would facilitate learning a debiased recommender system. Also, to ensure that the learned factors are independent of one another, we utilize Integral Probability Metric~\cite{muller1997integral}. To ensure the remaining bias from the confounders is eliminated, we utilize context-aware importance sampling weights~\cite{hassanpour2019learning}. The significant contributions of this work are:
\begin{itemize}[leftmargin=*]
    \item Investigating a novel setting of mitigating confounding bias in social recommender systems, %with social network information,
    \item Proposing to use a disentangling approach, D2Rec to identify and capture the different factors with specific roles in the causal inference formulation of recommender systems, and
    \item Demonstrating the effectiveness of the proposed framework on various real-world datasets with comparative analysis and ablation studies.
\end{itemize}
The remaining sections of the paper are categorized as follows. In Section~\ref{rw}, we introduce the related work that has been developed in debiasing recommender systems and how our method varies in comparison to them. In Section~\ref{prelims}, we present the technical preliminaries for the problem of posing a recommender system from a causal perspective. In Section~\ref{section:D2Rec}, we propose a causal disentanglement model with technical details. In Section~\ref{res}, we show the results on two real-world datasets and discuss observations on different baselines and the relevance of different components in our model. We conclude this paper in Section~\ref{conc}.
\section{Related Work}
\label{rw}
% Considerable research has been done over the past few years in recommendations aligned along three branches. 
\noindent\textbf{Disentangling user and item representations for recommendation.} This branch of research focuses on disentangling user and item latent features for better and explainable recommendations. To better understand user preferences and identify system defects, the authors of~\cite{liu2020explainable} proposed to generate explainable recommendations with the help of a framework that brings transparency in the representation learning process. The authors propose to discriminate information from different layers of graph convolutional networks. Another line of work focuses on disentangling latent user representations for news recommendations. The authors aim to disentangle and learn latent factors that influence a user to share a news article by leveraging a neighborhood routing algorithm~\cite{hu2020graph}. The authors of~\cite{qian2021intent} proposed a model to disentangling user and item latent representations for better recommendations. The user representations are disentangled into conformity influence and personal interest factors to improve recommendations of long-tail items. The item representations are disentangled the item attributes have a causal relation in the user preferences~\cite{nema2020untangle}, and disentangling user representations into user interest and conformity in an implicit feedback scenario~\cite{zheng2021disentangling}.

\noindent\textbf{Debiasing Recommender Systems.} Recommender systems suffer from various biases~\cite{chen2020bias}. As a result, various methods have been developed to facilitate debiasing the recommender systems to understand user preferences better. For instance, to mitigate selection bias, recent efforts include leveraging propensity score-based methods~\cite{schnabel2016recommendations}. Conformity bias is mitigated by leveraging user social relations~\cite{tang2012mtrust}. To deal with popularity bias, efforts leveraging disentanglement of latent representations to represent interest and conformity factors are used ~\cite{zheng2021disentangling}. Methods leveraging re-weighting techniques for unobserved samples with a uniform weight are developed to control exposure bias~\cite{yu2017selection}. Similarly, regularization-based frameworks are used to enhance long-tail coverage of items in a learning to rank algorithm~\cite{abdollahpouri2017controlling}.

\noindent\textbf{Causal Recommender Systems.} This branch of research aims to pose the recommender systems from a causal perspective. The earliest work in this field focuses on posing the recommendation problem as a causal inference problem rather than a prediction problem. In this case, the exposure is considered as treatment, and the ratings are considered as outcomes~\cite{liang2016causal}. This work was later extended by modeling the exposure and using it as a substitute for confounders~\cite{wang2020causal}. A recent work proposed in~\cite{li2021causal} utilized user's social relations to estimate the exposure along with propensity score and utilized this estimated exposure to mitigate selection bias. Another line of work aims to learn an optimal recommendation policy for each user. The authors aim to improve the recommender policy of the system by considering logged feedback and predict recommendation outcomes according to behavior under random exposure~\cite{bonner2018causal}. Some works, including~\cite{zhang2021causal} proposed to leverage the good aspects of popularity bias and deconfound the bad aspects for improving recommendations. Moreover, to improve the personalized rankings of recommender systems~\cite{yang2021top} proposed to apply Pearl's causal inference framework~\cite{peters2017elements}.

Compared to earlier causal recommendation works, our work differs in two ways. (1) Previous works utilize different estimation models to learn the substitute of the hidden confounders~\cite{liang2016causal,wang2020causal}, such as estimating exposure using the Poisson Factorization model and utilizing the estimated exposure as a substitute for the confounders. Although methods leveraging user's social relations to account for the confounders are proposed in~\cite{li2021causal} they do not consider any form of disentanglement. At the same time, they learn the user and item representations from the user-item networks. In contrast, our work aims to utilize disentanglement to learn a representation for the hidden confounders in the latent space based on all the observed information including the user's social connections and an item's popularity. (2) Works such as DICE~\cite{zheng2021disentangling} consider the causal perspective of recommender systems with implicit feedback, which leads to a different causal graph from the one considered in this work. For instance, unlike explicit feedback, the observed data for implicit feedback consists of a matrix of those items that the user viewed or clicked. Thus, DICE disentangles the causes of clicks into two components, i.e., user interest and item popularity. However, our work emphasizes the explicit feedback setting, which is more descriptive about the user preference as they include a rating score highlighting how much a user prefers an item. Moreover, we also include the effect induced by the user's social connections, which is not considered in DICE. This work studies debiasing recommender systems with explicit feedback from a causal perspective with network information.
%Thus, we believe we are the first to investigate recommender systems from a causal perspective by leveraging network information and a disentanglement framework for generating debiased recommendations.

\begin{table}[t]
    \centering
    %\footnotesize
    \resizebox{0.6\textwidth}{!}{\begin{tabular}{|c|c|}
    \hline
    \textbf{Symbol} & \textbf{Description} \\
    \hline
    $\bm{\theta_{u}}$ & Pre-trained user embedding from unsigned network\\
    \hline
    $\bm{\beta_{i}}$ & Pre-trained item embedding from unsigned network\\
    \hline
    $\bm{\alpha_{u}}$ & Disentangled user factor responsible for predicting exposure\\
    \hline
    $\bm{\alpha_{i}}$ & Disentangled item factor responsible for predicting exposure\\
    \hline
    $\bm{\gamma_{u}}$ & Disentangled user factor responsible for confounders\\
    \hline
    $\bm{\gamma_{i}}$ & Disentangled item factor responsible for confounders\\
    \hline
    $\bm{\Delta_{u}}$ & Disentangled user factor responsible for predicting rating\\
    \hline
    $\bm{\Delta_{i}}$ & Disentangled item factor responsible for predicting rating\\
    \hline
    %$\Upsilon_{u}$ & user coefficient for measuring contribution of confounders.\\
    %\hline
    $h_{i}$ & Feedforward neural networks\\
    \hline
    $\kappa$ & Coefficient that controls for discrepancy loss\\
    \hline
    $\sigma$ & Sigmoid activation function\\
    \hline
    \text{ReLU} & ReLU activation function\\
    \hline
    $y_{u,i}$ & Ground truth rating given to item $i$ by user $u$\\
    \hline
    $\hat{y}_{u,i}$ & Predicted rating that user $u$ would give to item $i$\\
    \hline
    $\exp_{u,i}$ & True exposure that user $u$ has interacted with item $i$\\
    \hline
    $\hat{\exp}_{u,i}$ & Predicted exposure that user $u$ would interact with item $i$\\
    \hline
    \end{tabular}}
    \caption{Notations.}
    \label{tab1}
    \vspace{-6mm}
\end{table}
\section{Preliminaries}
\label{prelims}
In this section, we first present the technical preliminaries. The notations used in this work are summarized in Table~\ref{tab1} where the vectors are represented in bold. Causal inference deals with the science of cause and effect. It helps understand the impact of performing interventions on the data. Traditional recommender systems~\cite{he2017neural,mnih2008probabilistic} are generally presented as a prediction problem, where, given the user and item latent representations, the model utilizes techniques such as matrix factorization to predict unseen ratings. However, these systems suffer from the Missing Not At Random (MNAR) problem~\cite{marlin2009collaborative} which may not lead to accurate predictions.
Posing the recommender system problem from a causal perspective can help in generating debiased recommendations. Earlier works~\cite{wang2020causal,liang2016causal} considered exposure as treatments and the observed ratings as the observed outcomes due to the treatment. One difficulty with the causal approach is the presence of confounders which are defined as the set of features that affect the treatment and outcomes. For instance, the mechanism with which a user is either exposed or not exposed to an item can influence the exposure value and the rating value for that user-item pair.
%\\
When the contribution due to the confounders is not included while modeling the rating predictions, it may lead to biased recommendations. A recent work, \cite{wang2020causal}, proposed a deconfounded recommender to explicitly control for confounding bias by accounting for the effects induced by the confounders.
A Poisson factorization model~\cite{gopalan2015scalable} is used for modelling the exposure as:
\begin{equation}
\exp_{u,i} | \bm{\pi}_{u}, \bm{\lambda}_{i} \sim Poisson(\bm{\pi}_{u}^{T} \bm{\lambda}_{i}),
\end{equation}
where $\bm{\pi}_u$ and $\bm{\lambda}_i$ are the user and item embeddings for modeling exposure.
Then, the deconfounded recommender controls the confounding bias by using the modelled exposure as a substitute for the hidden confounders as follows:
\begin{equation}
    y_{u,i} = \bm{\theta}_u^T\bm{\beta}_i \cdot \exp_{u,i} + {\gamma}_u\hat{\exp}_{u,i} + \epsilon_{u,i},
    \label{eq:deconf_recsys}
\end{equation}
where $\epsilon_{u,i} \sim \mathcal{N}(0,\sigma^2)$ is the noise term.
The term $\bm{\theta}_u^T\bm{\beta}_i \cdot \exp_{u,i}$ models the rating prediction for which user $u$ was exposed to the item $i$. ${\gamma}_u\hat{\exp}_{u,i}$ models the effect of the substitute learned for the hidden confounders, on the outcome, where ${\gamma}_u$ is a user-specific coefficient that represents the sign and the magnitude of this effect and $\hat{\exp}_{u,i}$ represents the substitute for the hidden confounders, i.e., the inferred (probability of) exposure.
When $\gamma_u$ is positive, in Eq.~\eqref{eq:deconf_recsys}, the product ${\gamma}_u\hat{\exp}_{u,i}$ models the positive correlation between the confounder and rating.
\section{Causal Disentanglement for Debiased Recommender (D2Rec)}
\label{section:D2Rec}

% \begin{algorithm}
% \SetAlgoLined
% \algorithmicrequire{ $\theta_{u}$ and $\beta_{i}$ obtained from equations Eq. \eqref{eq5} and Eq. \eqref{eq6} respectively;}
% %\\
%  \While{not converge}{
%   sample batch of users $U \sim p(u)$\;
%   sample batch of items $I \sim p(i)$\;
%   \For{user u in U}
%   {
%     obtain disentangled representations $\alpha_{u}$,$\gamma_{u}$ and $\Delta_{u}$ from equation \ref{disentangler}. 
%   }
%   \For{item i in I}
%   {
%     obtain disentangled representations $\alpha_{i}$,$\gamma_{i}$ and $\Delta_{i}$ from equation \ref{disentangler}. 
%   }
%   Compute joint disentangled representation for $u$ and $i$ from Eq. \eqref{merger};\\
%   Obtain $\hat{exp}_{u,i}$ from equation Eq. \eqref{exp} for all users and items in the batch;\\
%   Obtain $\hat{y}_{u,i}$ from equation Eq. \eqref{rating} for all users and items in the batch;\\
%   Evaluate Eq. \eqref{exp_loss} and Eq. \eqref{rate_loss};\\
%   Update $\alpha_{u}$,$\gamma_{u}$,$\Delta_{u}$,$\alpha_{i}$,$\gamma_{i}$ and $\Delta_{i}$ and $\Upsilon_{u}$;
%  }
%  \caption{Disentangled and debiased Social Recommender Algorithm}
% \end{algorithm}

This work aims to debias the social recommender systems from a causal perspective by leveraging auxiliary network information, including the user-social network and user-item interaction network. The goal is to learn three independent disentangled factors from the user and item's latent representations and control for the confounding bias. In the proposed \textit{Causal Disentanglement for DeBiased Recommendations (D2Rec)} model, the user representation learned from the user-social network acts as a diverse source of information that consists of factors for the user's preferences and factors for the hidden confounders (e.g., conformity). Similarly, item representations consist of the factors for the item attributes and the factors for the hidden confounders (e.g., item popularity). Thus, D2Rec aids in causally disentangling these learned representations to account for the hidden confounders, the factors for causing the exposure (treatment) and the factors for causing the rating (outcome). Furthermore, to ensure each factor is independent of the other factors, we utilize discrepancy loss~\cite{gretton2012kernel,muller1997integral} to maximize the distance between the distribution of the learned disentangled factors, similar to~\cite{zheng2021disentangling}.

\begin{figure}[h]
    \centering
    \includegraphics[width=0.7\textwidth]{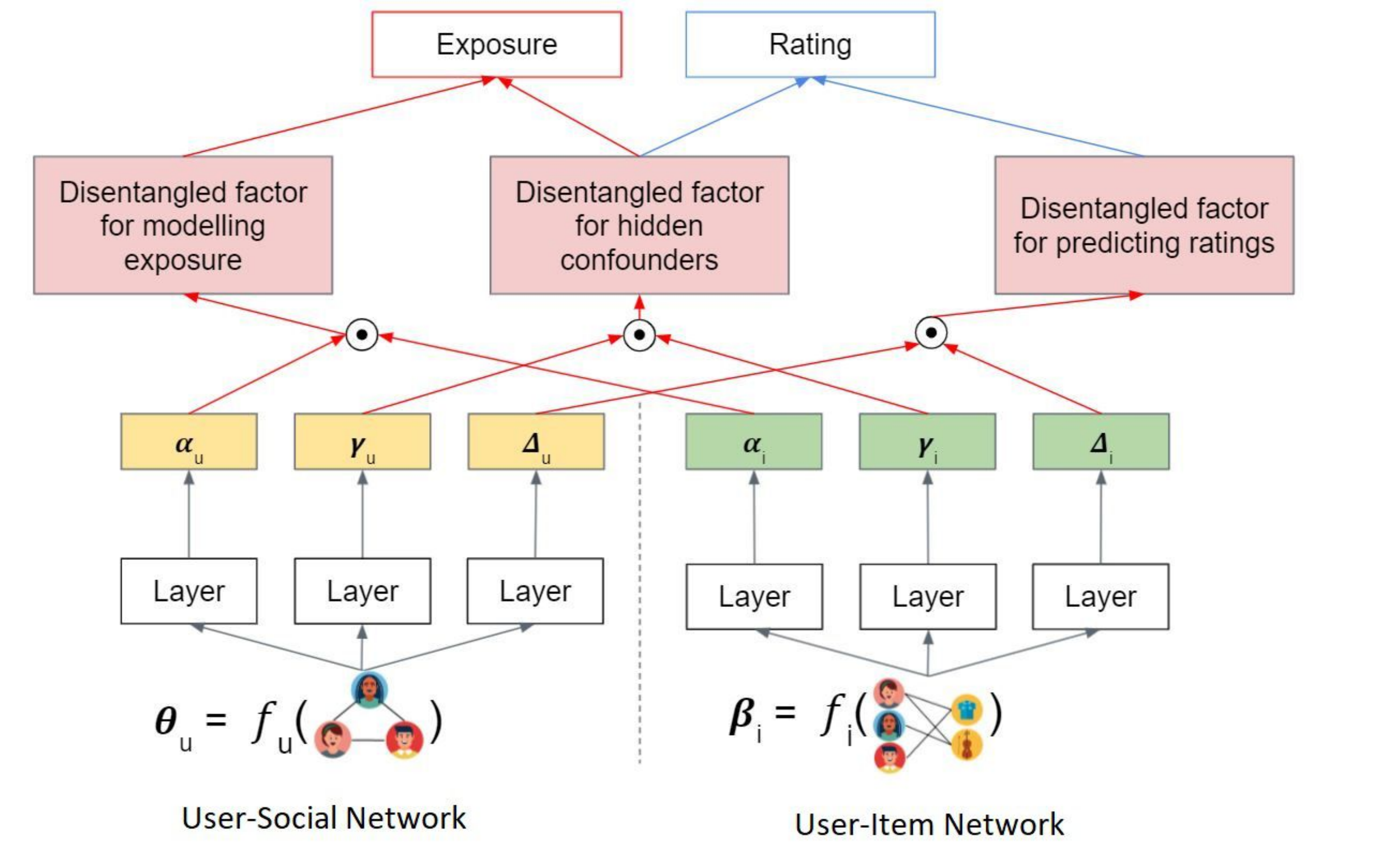}
    \caption{An overview of the architecture of the proposed D2Rec model. Here,
    $f_{u}$ represents the function to learn user embeddings from the user-social network, $f_{i}$ represents the function to learn item embeddings from the user-item interactions network, $\bm{\theta_{u}}$ and $\bm{\beta_{i}}$ represents the user embeddings and item embeddings, Layer represents a feedforward neural network used to learn the disentangled factors. $\bm{\alpha_{u}}$,$\bm{\alpha_{i}}$,$\bm{\gamma_{u}}$,$\bm{\gamma_{i}}$,$\bm{\Delta_{u}}$,$\bm{\Delta_{i}}$ represent the learned disentangled user and item factors.}
    \label{fig1}
    \vspace{-5mm}
\end{figure}

An overview of the proposed D2Rec can be found in Fig.~\ref{fig1}. The approach consists of three key components. First, it has a node embedding learning module 
%based on Node2vec~\cite{grover2016node2vec} 
that takes (1) the user-social network as input to learn the user embeddings; and (2) the user-item interaction network as input for learning the item embeddings. 
%More details about this module can be found in the experimental settings. 
Second, we propose a novel module namely, \textit{causally disentangled representation learning networks} that disentangles the previously obtained user and item embeddings into three factors that affect both the exposure (treatment) and the ratings (outcomes). The third component of the proposed model is a neural network that takes the factors mentioned above as the input and predicts the rating and the exposure of a given user-item pair.
\subsection{Representation Learning Networks for Causal Disentanglement}
Recent efforts, such as~\cite{hassanpour2019learning}, provide evidence that observational data can be disentangled into its corresponding underlying factors, which can improve performance against the downstream tasks. The disentangled factors facilitate in a better understanding of the information present in the latent space. As the user representations and item representations are collective representations of multiple factors, the proposed disentangling component can better capture different factors of the user and item representations. Since we aim to look at recommender systems from a causal perspective, we propose to disentangle the user and item latent representations $(\bm{\theta_{u}}$ and $\bm{\beta_{i}})$ into three underlying factors {$\alpha$, $\gamma$, $\Delta$}. Among them, $\alpha$ is partially responsible for modeling the exposure (treatment), the factor $\Delta$ is partially responsible for predicting the ratings (outcomes), and $\gamma$ is the factor responsible for the confounding factors that causally affect both the exposure and the ratings.
%\\
We use the user-social network to learn the embedding for each user, and we use the user-item interaction network to learn the embedding for each item. Note that we only use the user-social network to learn the user embeddings. It is possible to learn user embeddings by leveraging both user-social networks and user-item interaction networks~\cite{fan2019graph}, which we leave for future work. The user and item embeddings are defined as:
\begin{gather}
\bm{\theta_{u}} = \max _{f} \sum_{u \in V} \log \operatorname{Pr}\left(N_{S}(u) \mid f(u)\right),\quad
\bm{\beta_{i}} = \max _{f} \sum_{i \in V} \log \operatorname{Pr}\left(N_{S}(i) \mid f(i)\right).
\label{eq5}
\end{gather}
The representation networks consist of six independent layers that facilitate the learning of the disentangled factors from the obtained user representations and item representations $\bm{\theta_{u}}$ and $\bm{\beta_{i}}$. They are $\bm{\alpha_u}$, $\bm{\alpha_i}$, $\bm{\gamma_u}$, $\bm{\gamma_i}$, $\bm{\Delta_u}$, and $\bm{\Delta_i}$, denoting the underlying factors for exposure prediction ($\alpha$), confounders ($\gamma$), and rating prediction  ($\Delta$). Formally,
\begin{equation}
\begin{aligned}
    \bm{\alpha_{u}} = \text{ReLU}(h_{1}(\bm{\theta_{u}})) ;\quad \bm{\alpha_{i}} = \text{ReLU}(h_{4}(\bm{\beta_{i}}));\\
    \bm{\gamma_{u}} = \text{ReLU}(h_{2}(\bm{\theta_{u}})) ;\quad \bm{\gamma_{i}} = \text{ReLU}(h_{5}(\bm{\beta_{i}}));\\
    \bm{\Delta_{u}} = \text{ReLU}(h_{3}(\bm{\theta_{u}})) ;\quad \bm{\Delta_{i}} = \text{ReLU}(h_{6}(\bm{\beta_{i}})).
\end{aligned}
\label{disentangler}
\end{equation}
where ReLU represents the nonlinear ReLU activation function, $h_{k}, k = 1,...,6$ denote feedforward neural networks.

\subsection{Rating and Exposure Prediction}
We leverage the disentangled factors for users and items obtained from Eq.~\eqref{disentangler} for rating prediction and exposure modeling for user-item pairs;
each factor plays a pivotal role in either computing the rating or the exposure. To ensure the learned representations are independent of one another, we use an Integral Probability Metric -- Maximum Mean Discrepancy ($\operatorname{MMD}$)~\cite{gretton2012kernel}, which aims to measure how close two distributions are. The discrepancy loss~\cite{hassanpour2019learning} is given by,
\begin{equation}
%\small
\begin{aligned}
 \mathcal{L}_{disc} = \operatorname{MMD}(\bm{\alpha_u},\bm{\gamma_u}) + \operatorname{MMD}(\bm{\alpha_u},\bm{\Delta_u}) + \operatorname{MMD}(\bm{\gamma_u},\bm{\Delta_u}) +
 \operatorname{MMD}(\bm{\alpha_i},\bm{\gamma_i}) +
 \operatorname{MMD}(\bm{\gamma_i},\bm{\Delta_i}) +
 \operatorname{MMD}(\bm{\alpha_i},\bm{\Delta_i}), 
\end{aligned}
\label{dicrep}
\end{equation}
in which MMD refers to,
\begin{equation}
%\small
\operatorname{MMD}(\mathbb{P}, \mathbb{Q} ; \mathcal{F}):=\sup _{f \in \mathcal{F}}|\mathbb{E}[f(X)]-\mathbb{E}[f(Y)]|.
\end{equation}
Here, $\mathbb{P}$ and $\mathbb{Q}$ are probability distributions, and $\mathcal{F}$ is a set containing all continuous functions.
We take each disentangled user factor and perform a Hadamard product ($\odot$) to its counterpart in the item factors to generate a joint user-item representation for the particular disentangled factor. We obtain the combined factors, $\alpha_{u,i}$, $\gamma_{u,i}$, and $\Delta_{u,i}$ as follows:
\begin{equation}
\begin{aligned}
    \alpha_{u,i} = \bm{\alpha_{u}} \odot \bm{\alpha_{i}};\quad
    \gamma_{u,i} = \bm{\gamma_{u}} \odot \bm{\gamma_{i}};\quad
    \Delta_{u,i} = \bm{\Delta_{u}} \odot \bm{\Delta_{i}}.
\end{aligned}
\label{merger}
\end{equation}
Looking from a causal perspective, the factor $\alpha_{u,i}$ is used for modeling the treatment (exposure), the factor $\gamma_{u,i}$ is used for modeling the confounders that affect both the treatment (exposure) and the outcome (ratings), and the factor $\Delta_{u, i}$ is used for modeling the outcomes (ratings). We then group $\alpha_{u,i}$, $\gamma_{u,i}$, and $\Delta_{u,i}$ into the following categories: the factors $\alpha_{u,i}$ and $\gamma_{u,i}$ are collectively used for modelling the exposure, and factors $\gamma_{u,i}$ and $\Delta_{u,i}$ are collectively used for modelling the ratings.
Traditional recommenders utilize the entangled user and item representations to minimize the errors on the observed ratings. However, based on Fig.~\ref{causal_graph} we argue that not all components of the representations are helpful for rating prediction. Thus, with the aid of disentanglement, we learn factors from the user and item latent representations that only affect the ratings and are independent of the exposure mechanism and vice versa (since $\Delta_{u,i}$ and $\alpha_{u,i}$ are independent of each other) which helps the rating prediction model to generate debiased recommendations. Given the disentanglement, we proceed to compute the exposure as %Also, by explicitly accounting for confounders as inputs to the rating prediction we can adjust for the confounding bias adjusting for confounding bias by taking confounders as its input. 
\begin{equation}
 \hat{\exp}_{u,i} = \sigma(\alpha_{u,i} \cdot \gamma_{u,i}),
\label{exp}
\end{equation}
where $\cdot$ represents the dot product, $\sigma$ is the sigmoid activation function. Similarly, the rating can be computed as
%and $\Upsilon_{u}$ measures the contribution of the confounding factor $\gamma_{u,i}$ in modelling the exposure. 
\begin{equation}
 \hat{y}_{u,i} = \omega(\exp_{u,i}, \hat{\exp}_{u,i}, \gamma_{u,i}) \cdot \text{ReLU}(\gamma_{u,i} \cdot \Delta_{u,i}),
 \label{rating}
\end{equation}
where $\exp_{u,i}$ denotes the true exposure for user $u$ and item $i$, $\hat{\exp}_{u,i}$ represents the predicted exposure learned from Eq.~\eqref{exp} and $\omega(\exp_{u,i}, \hat{\exp}_{u,i}, \gamma_{u,i})$ is the re-weighting function. By taking the confounding factors as an input to the rating prediction function we mitigate the confounding bias.
%\\
As mentioned earlier, the factor $\gamma$ is likely to exhibit some degree of bias, as it consists of the information affecting both the exposure and ratings. To counter this, we rely on \textit{context-aware} importance sampling weights that could mitigate the remainder bias~\cite{hassanpour2019learning}:
\begin{equation}
%\small
\begin{aligned}
    \omega&(\exp_{u,i}, \hat{\exp_{u,i}}, \gamma_{u,i}) &=1+\frac{\operatorname{Pr}\left(\exp_{u,i}=1\right)}{1-\operatorname{Pr}\left(\exp_{u,i}=1\right)} \times \frac{1-\operatorname{Pr}\left(\hat{\exp}_{u,i}=1 \mid \gamma_{u,i}\right)}{\operatorname{Pr}\left(\hat{\exp}_{u,i}=1 \mid \gamma_{u,i}\right)},
\end{aligned}
\
\end{equation}
where $\exp_{u,i}$ represents the true exposure, i.e. whether the user was exposed to the item and $\hat{\exp}_{u,i}$ represents the predicted exposure learned from Eq.~\eqref{exp}. The reweighting function is effective as no confounders are discarded, and only the legitimate confounders are used to derive the weights. Once we obtain the predicted exposure from Eq.~\eqref{exp} and predicted ratings from Eq.~\eqref{rating}, we compute the overall objective function as
\begin{equation}
\begin{aligned}
 J(\hat{y}_{u,i},\hat{\exp}_{u,i}) =  \kappa \cdot \mathcal{L}_{disc} + \mathcal{L}(y_{u,i},\hat{y}_{u,i}) +
\mathcal{L}_{\exp}(\exp_{u,i},\hat{\exp}_{u,i}),
\end{aligned}
\end{equation}
where we aim to minimize $\mathcal{L}(y_{u,i},\hat{y}_{u,i})$, which represents the mean squared error in predicting the ratings:
\begin{equation}
    \mathcal{L}=\sum_{u,i}(y_{u,i}-\hat{y}_{u,i})^{2}.
\label{rate_loss}
\end{equation}
We also aim to minimize $\mathcal{L}_{\exp}(\exp_{u,i},\hat{\exp}_{u,i})$, which represents the binary cross-entropy loss formulated as
\begin{equation}
\begin{split}
    \mathcal{L}_{\exp}= -\sum_{u,i}(\exp_{u,i} \cdot  \log (\hat{\exp}_{u,i}) +(1-\exp_{u,i})\cdot\log(1-\hat{\exp}_{u,i})).
\label{exp_loss}
\end{split}
\end{equation}
To ensure that the learned disentangled representations are independent of one another, we use the discrepancy loss in Eq.~\eqref{dicrep} that measures how close two distributions are. Therefore, the goal is to maximize this discrepancy loss. Furthermore, since the discrepancy loss directly influences the distribution of the learned embeddings, we use $\kappa$ to control the effect of discrepancy loss on the overall objective function.
\section{Experiments}
\label{res}
The node embedding module of D2Rec learns the user and item embeddings from social network information. We leverage the Node2vec~\cite{grover2016node2vec} framework to obtain the user and item embeddings. 
%The Node2vec framework uses a flexible biased random walk procedure that can explore neighborhoods in a Breadth-First Search (BFS) as well as Depth-First Search (DFS) fashion. 
More advanced node embedding methods, as well as GNN based methods to directly incorporate network information while learning the user and item representations can also be used with D2Rec and, will be explored in future work.

We conducted a series of experiments to understand whether disentangling the user and item latent features learned from auxiliary network information can help adjust for confounding bias in debiased social recommendations. Ideally, a causal method is evaluated based on how well an algorithm mitigates confounding bias with a test set where treatments are randomly assigned~\cite{shadish2008can}.
%As we formulate social recommender systems with explicit feedback as a causal inference problem, it is expected to perform well in an unbiased test set where treatments (exposure) are assigned through randomized controlled trials~\cite{shadish2008can}. Randomized controlled trials assign items to the users at random, mitigating any effect of the confounders. Datasets with randomized test sets such as Yahoo!R3 and Coat~\cite{wang2020causal,schnabel2016recommendations} are widely used to evaluate causal recommender algorithms.

Since the existing real-world social recommender system datasets do not have unbiased test sets, we need to first create pseudo unbiased test sets through sampling from observational data. We follow the standard protocol introduced by~\cite{liang2016causal,bonner2018causal} to create such pseudo unbiased test sets. In a pseudo unbiased test set, items are uniformly exposed, i.e., each item has precisely the same number of appearances in the test set. In particular, we split each dataset into training and test sets as follows. First, the training samples are randomly sampled from the original data (thus biased). Then, from the rest of the dataset, we create subsets as the unbiased test sets by conditioning on item popularity to make each pseudo unbiased test set have an equal number of ratings for items. In this way, we ensure that the pseudo unbiased test set has different exposure distribution from its corresponding training set. Thus, we can verify whether D2Rec effectively adjusts for the confounding bias by computing its generalization performance on the pseudo unbiased test sets~\cite{buhlmann2020invariance}. We %generate test sets ranging from 2 through 10 ratings per item and 
conduct experiments to answer the following research questions.
\begin{itemize}[leftmargin=*]
\item \textbf{RQ.1} Can disentangling the user and item embeddings with network information %facilitate learning user preference, and therefore,
help debias recommendations?
%\item \textbf{RQ.2} Does network information play a pivotal role in understanding true user preferences?
%\item \textbf{RQ.3} Does the disentangled factor, $\gamma$, account for the confounders from the network information?
\item \textbf{RQ.2} What are the roles played by the network information and by the disentanglement module concerning the performance of D2Rec, respectively?
\end{itemize}%
% Our evaluation strategy is based on the observation that {if D2Rec can outperform the state-of-the-art baselines in causal recommendation systems on such pseudo unbiased test sets, then we can justify that the disentangling factors facilitate in capturing true user and item features which leads to improved generalization performance.}
\subsection{Experiment Setup}
This section introduces the datasets, evaluation metrics, parameter settings, and the baselines used for the experiments. To simulate the causal evaluation settings as mentioned earlier, we generate test sets ranging from 2 through 10 ratings per item. Finally, we report the performance comparison across the different baselines and perform an ablation study on various components of D2Rec to comment on their significance.
\subsubsection{Datasets}
For our experiments, we use two representative real-world datasets Ciao\footnote{\url{https://www.cse.msu.edu/~tangjili/Ciao.rar}} and Epinions\footnote{\url{https://www.cse.msu.edu/~tangjili/Epinions.rar}}. Both these datasets are derived from popular social networking websites Ciao\footnote{http://www.ciao.co.uk} and Epinions\footnote{www.epinions.com}. These websites allow users to rate multiple items, browse/write reviews, and formulate trust/distrust relations among users. They can provide a vast amount of rating information and social information. The social interactions denote the trust relationships between users. We created self-loops for users who did not have any social interactions denoting that they trusted themselves. The ratings are in the range of 1 to 5. A summary of the datasets can be found in Table~\ref{table1}.
\begin{table}[]
\centering
\resizebox{\textwidth}{!}{%
\begin{tabular}{|c|c|c|c|c|c|c|}
\hline
\textbf{Dataset} & \textbf{No. of Users} & \textbf{No. of Items} & \textbf{No. of Ratings} & \textbf{Rating Density} & \textbf{No. of Social Connections} & \textbf{Social Connections Density} \\ \hline
\textbf{Ciao} & 7,375 & 105,114 & 284,086 & 0.0366 & 112,384 & 0.2060 \\ \hline
\textbf{Epinions} & 40,163 & 139,738 & 664,824 & 0.0118 & 455,751 & 0.0282 \\ \hline
\end{tabular}%
}
\caption{Statistics of the Datasets.}
\label{table1}
\vspace{-8mm}
\end{table}
% \begin{table}
% \centering
% \small
% \begin{tabular}{|c|c|c|}
% \hline
% Dataset& Ciao & Epinions \\
% \hline
% No. of Users&7,375&40,163 \\
% \hline
% No. of Items&105,114&139,738 \\
% \hline
% No. of of Ratings&284,086&664,824 \\
% \hline
% Rating Density&0.0366&0.0118 \\
% \hline
% No. of of Social Connections&112,384&455,751 \\
% \hline
% Social Connections Density &0.2060&0.0282 \\
% \hline
% \end{tabular}
% \caption{Statistics of the Datasets}
% \label{table1}
% \vspace{-10mm}
% \end{table}
\subsubsection{Evaluation Metrics} For evaluation of D2Rec and baselines, we rely on rating prediction and ranking metrics.

\noindent{\bf Rating Prediction Metrics.} To evaluate rating performance, we use two popular metrics, Mean Squared Error (MSE) and Mean Absolute Error (MAE):
%The MAE can be formulated as
%and NDCG@10 and Hit-Ratio@10 for ranking performance. 
\begin{equation}
\begin{aligned}
    MAE &=& \frac{1}{N_{ts}} \sum_{(u,i)} |y_{u,i} - \hat{y}_{u,i}|,\quad
    MSE &=& \frac{1}{N_{ts}} \sum_{(u,i)} (y_{u,i} - \hat{y}_{u,i})^2.\label{eq11}
\end{aligned}
%\label{disentangler}
\end{equation}
% \begin{eqnarray}
%     MAE &=& \frac{1}{N_{ts}} \sum_{(u,i)} |y_{u,i} - \hat{y}_{u,i}|,\\\label{eq10}
% %\end{equation}
% %The MSE is given by,
% %\begin{equation}
%     MSE &=& \frac{1}{N_{ts}} \sum_{(u,i)} (y_{u,i} - \hat{y}_{u,i})^2.\label{eq11}
%\end{eqnarray}
Here $N_{ts}$ denotes the total number of ratings in the test set, $y_{u,i}$ denotes the true observed rating for the entry of the user-item pair $(u,i)$, and $\hat{y}_{u,i}$ denotes the predicted rating for the corresponding user-item pair.

\noindent{\bf Ranking Metrics.} To evaluate personalized ranking performance, we use two popular metrics, Hit-Ratio@K and Normalized Discounted Cumulative Gain (NDCG@K):
\begin{itemize}[leftmargin=*]
    \item Hit-Ratio@K refers to the proportion of users for which the model can correctly include the items a user has interacted within the list of top-K recommended items 
    %$HR=\frac{|U_{\text {hit }}^{K}|}{|U_{\text {all}}|}$
%It can be formulated as,
 \begin{equation}
 HR=\frac{|U_{\text {hit }}^{K}|}{|U_{\text {all}}|},
 \end{equation}
where $|U_{\text {hit }}^{K}|$ is the number of users for which the recommender systems was able to include the items user has interacted with in the top-K recommended items list and $|U_{\text {all}}|$ denotes the total number of users in the test set.
\item For user-item interactions, gain for an item refers to the relevance score. To take order of the ranking into consideration, Discounted Cumulative Gain (DCG) is formulated as
\begin{equation}
DCG_{K}^{i}=\sum_{r=1}^{K} \frac{2^{{y}_{u,r}}-1}{\log _{2}(1+r)},
\end{equation}
where ${y}_{u,r}$ refers to the ground truth rating for user $u$ and $r^{th}$ ranked item. The normalized discounted gain (nDCG) is then defined as 
\begin{equation}
nDCG_{K}=\frac{1}{|U|} \sum_{u \in U} \frac{DCG_{K}^{u}}{IDCG_{K}^{u}},
\end{equation}
where $|U|$ refer to the total number of users in the test set and $IDCG_{K}^{u}$ refers to the best possible value of $DCG_{K}^{u}$.
\end{itemize}
\subsubsection{Parameter Settings}
We implemented all baseline models on a Linux server with Tesla K-80 25GB GPU. We implemented D2Rec in Pytorch. To verify if D2Rec is helping with debiasing recommender systems, we first perform a train/test split of 60/40; we further split the test set into subgroups conditioned in terms of popularity to obtain subsets of items with 2 through 10 ratings per item. The embedding size, $d$, was varied from \{32,64,128,256\}. We varied the batch size from \{64,128,512,1000\} and the learning rate was varied from \{0.0001,0.001,0.01\}. Moreover, we set the activation function as ReLU. We employed six different neural networks for calculating $\bm{\alpha_{i}}$, $\bm{\alpha_{u}}$, $\bm{\gamma_{i}}$, $\bm{\gamma_{u}}$, $\bm{\Delta_{i}}$, and $\bm{\Delta_{u}}$. The maximum number of epochs was set to 200, and an early stopping strategy was performed, where we stopped training if the MSE on the train set did not decrease for ten successive epochs. The parameters were initialized with the corresponding papers' values and then tuned to achieve the best empirical performance for all baselines. We optimized all models with the Adam optimizer. $\kappa$ is set to 0.5 to denote the contribution of discrepancy loss to the objective functions.
% \usepackage{multirow}
% \usepackage{graphicx}
\begin{table}[]
\centering
%\small
\resizebox{0.85\textwidth}{!}{%
\begin{tabular}{|c|cccccccc|}
\hline
\multirow{3}{*}{\textbf{Model}} & \multicolumn{8}{c|}{\textbf{Popularity Debiased Test Sets}} \\ \cline{2-9} 
 & \multicolumn{2}{c|}{\textbf{Popularity=2}} & \multicolumn{2}{c|}{\textbf{Popularity=3}} & \multicolumn{2}{c|}{\textbf{Popularity=5}} & \multicolumn{2}{c|}{\textbf{Popularity=10}} \\ \cline{2-9} 
 & \multicolumn{1}{c|}{\textbf{NDCG@10}} & \multicolumn{1}{c|}{\textbf{HR@10}} & \multicolumn{1}{c|}{\textbf{NDCG@10}} & \multicolumn{1}{c|}{\textbf{HR@10}} & \multicolumn{1}{c|}{\textbf{NDCG@10}} & \multicolumn{1}{c|}{\textbf{HR@10}} & \multicolumn{1}{c|}{\textbf{NDCG@10}} & \textbf{HR@10} \\ \hline
\textbf{SocialMF~\cite{jamali2010matrix}} & \multicolumn{1}{c|}{0.32} & \multicolumn{1}{c|}{0.64} & \multicolumn{1}{c|}{0.34} & \multicolumn{1}{c|}{0.67} & \multicolumn{1}{c|}{0.35} & \multicolumn{1}{c|}{0.70} & \multicolumn{1}{c|}{0.37} & 0.73 \\ \hline
\textbf{GraphRec~\cite{fan2019graph}} & \multicolumn{1}{c|}{0.24} & \multicolumn{1}{c|}{0.56} & \multicolumn{1}{c|}{0.23} & \multicolumn{1}{c|}{0.55} & \multicolumn{1}{c|}{0.20} & \multicolumn{1}{c|}{0.52} & \multicolumn{1}{c|}{0.20} & 0.51 \\ \hline
\textbf{ConsisRec~\cite{yang2021consisrec}} & \multicolumn{1}{c|}{0.36} & \multicolumn{1}{c|}{0.70} & \multicolumn{1}{c|}{0.38} & \multicolumn{1}{c|}{0.74} & \multicolumn{1}{c|}{0.41} & \multicolumn{1}{c|}{0.77} & \multicolumn{1}{c|}{0.43} & 0.79 \\ \hline
\textbf{NeuMF~\cite{he2017neural}} & \multicolumn{1}{c|}{0.31} & \multicolumn{1}{c|}{0.65} & \multicolumn{1}{c|}{0.31} & \multicolumn{1}{c|}{0.66} & \multicolumn{1}{c|}{0.32} & \multicolumn{1}{c|}{0.67} & \multicolumn{1}{c|}{0.33} & 0.69 \\ \hline
\textbf{PMF~\cite{mnih2008probabilistic}} & \multicolumn{1}{c|}{0.33} & \multicolumn{1}{c|}{0.69} & \multicolumn{1}{c|}{0.34} & \multicolumn{1}{c|}{0.71} & \multicolumn{1}{c|}{0.36} & \multicolumn{1}{c|}{0.72} & \multicolumn{1}{c|}{0.38} & 0.74 \\ \hline
\textbf{IPS-MF~\cite{liang2016causal}} & \multicolumn{1}{c|}{0.31} & \multicolumn{1}{c|}{0.65} & \multicolumn{1}{c|}{0.32} & \multicolumn{1}{c|}{0.68} & \multicolumn{1}{c|}{0.38} & \multicolumn{1}{c|}{0.73} & \multicolumn{1}{c|}{0.40} & 0.78 \\ \hline
\textbf{CIRS~\cite{wang2020causal}} & \multicolumn{1}{c|}{0.37} & \multicolumn{1}{c|}{0.70} & \multicolumn{1}{c|}{\textbf{0.39}} & \multicolumn{1}{c|}{0.74} & \multicolumn{1}{c|}{0.39} & \multicolumn{1}{c|}{0.75} & \multicolumn{1}{c|}{0.42} & 0.80 \\ \hline
\textbf{DICE~\cite{zheng2021disentangling}} & \multicolumn{1}{c|}{0.34} & \multicolumn{1}{c|}{0.68} & \multicolumn{1}{c|}{0.36} & \multicolumn{1}{c|}{0.70} & \multicolumn{1}{c|}{0.38} & \multicolumn{1}{c|}{0.74} & \multicolumn{1}{c|}{0.42} & 0.81 \\ \hline
\textbf{D2Rec} & \multicolumn{1}{c|}{\textbf{0.38}} & \multicolumn{1}{c|}{\textbf{0.73}} & \multicolumn{1}{c|}{0.38} & \multicolumn{1}{c|}{\textbf{0.76}} & \multicolumn{1}{c|}{\textbf{0.41}} & \multicolumn{1}{c|}{\textbf{0.80}} & \multicolumn{1}{c|}{\textbf{0.44}} & \textbf{0.83} \\ \hline
\end{tabular}%
}
\caption{Comparing the ranking performance of different recommender system models with D2Rec for Epinions. }%Each test set is debiased by fixing the number of ratings per item.}
\label{table4}
\vspace{-5mm}
\end{table}

% \usepackage{multirow}
% \usepackage{graphicx}
\begin{table}[]
\centering
%\small
\resizebox{0.85\textwidth}{!}{%
\begin{tabular}{|c|cccccccc|}
\hline
\multirow{3}{*}{\textbf{Model}} & \multicolumn{8}{c|}{\textbf{Popularity Debiased Test Sets}} \\ \cline{2-9} 
 & \multicolumn{2}{c|}{\textbf{Popularity=2}} & \multicolumn{2}{c|}{\textbf{Popularity=3}} & \multicolumn{2}{c|}{\textbf{Popularity=5}} & \multicolumn{2}{c|}{\textbf{Popularity=10}} \\ \cline{2-9} 
 & \multicolumn{1}{c|}{\textbf{NDCG@10}} & \multicolumn{1}{c|}{\textbf{HR@10}} & \multicolumn{1}{c|}{\textbf{NDCG@10}} & \multicolumn{1}{c|}{\textbf{HR@10}} & \multicolumn{1}{c|}{\textbf{NDCG@10}} & \multicolumn{1}{c|}{\textbf{HR@10}} & \multicolumn{1}{c|}{\textbf{NDCG@10}} & \textbf{HR@10} \\ \hline
\textbf{SocialMF~\cite{jamali2010matrix}} & \multicolumn{1}{c|}{0.31} & \multicolumn{1}{c|}{0.60} & \multicolumn{1}{c|}{0.36} & \multicolumn{1}{c|}{0.70} & \multicolumn{1}{c|}{0.40} & \multicolumn{1}{c|}{0.77} & \multicolumn{1}{c|}{0.44} & 0.82 \\ \hline
\textbf{GraphRec~\cite{fan2019graph}} & \multicolumn{1}{c|}{0.16} & \multicolumn{1}{c|}{0.37} & \multicolumn{1}{c|}{0.14} & \multicolumn{1}{c|}{0.35} & \multicolumn{1}{c|}{0.15} & \multicolumn{1}{c|}{0.38} & \multicolumn{1}{c|}{0.18} & 0.44 \\ \hline
\textbf{ConsisRec~\cite{yang2021consisrec}} & \multicolumn{1}{c|}{0.29} & \multicolumn{1}{c|}{0.59} & \multicolumn{1}{c|}{0.32} & \multicolumn{1}{c|}{0.63} & \multicolumn{1}{c|}{0.35} & \multicolumn{1}{c|}{0.68} & \multicolumn{1}{c|}{0.37} & 0.72 \\ \hline
\textbf{NeuMF~\cite{he2017neural}} & \multicolumn{1}{c|}{0.26} & \multicolumn{1}{c|}{0.54} & \multicolumn{1}{c|}{0.28} & \multicolumn{1}{c|}{0.58} & \multicolumn{1}{c|}{0.30} & \multicolumn{1}{c|}{0.61} & \multicolumn{1}{c|}{0.32} & 0.68 \\ \hline
\textbf{PMF~\cite{mnih2008probabilistic}} & \multicolumn{1}{c|}{0.29} & \multicolumn{1}{c|}{0.60} & \multicolumn{1}{c|}{0.32} & \multicolumn{1}{c|}{0.66} & \multicolumn{1}{c|}{0.36} & \multicolumn{1}{c|}{0.70} & \multicolumn{1}{c|}{0.43} & 0.82 \\ \hline
\textbf{IPS-MF~\cite{liang2016causal}} & \multicolumn{1}{c|}{0.27} & \multicolumn{1}{c|}{0.55} & \multicolumn{1}{c|}{0.30} & \multicolumn{1}{c|}{0.60} & \multicolumn{1}{c|}{0.39} & \multicolumn{1}{c|}{0.74} & \multicolumn{1}{c|}{0.44} & 0.83 \\ \hline
\textbf{CIRS~\cite{wang2020causal}} & \multicolumn{1}{c|}{\textbf{0.34}} & \multicolumn{1}{c|}{\textbf{0.62}} & \multicolumn{1}{c|}{0.36} & \multicolumn{1}{c|}{0.68} & \multicolumn{1}{c|}{0.38} & \multicolumn{1}{c|}{0.73} & \multicolumn{1}{c|}{0.43} & 0.81 \\ \hline
\textbf{DICE~\cite{zheng2021disentangling}} & \multicolumn{1}{c|}{0.29} & \multicolumn{1}{c|}{0.58} & \multicolumn{1}{c|}{0.34} & \multicolumn{1}{c|}{0.66} & \multicolumn{1}{c|}{0.37} & \multicolumn{1}{c|}{0.70} & \multicolumn{1}{c|}{0.46} & 0.80 \\ \hline
\textbf{D2Rec} & \multicolumn{1}{c|}{0.30} & \multicolumn{1}{c|}{0.60} & \multicolumn{1}{c|}{\textbf{0.36}} & \multicolumn{1}{c|}{\textbf{0.70}} & \multicolumn{1}{c|}{\textbf{0.41}} & \multicolumn{1}{c|}{\textbf{0.79}} & \multicolumn{1}{c|}{\textbf{0.49}} & \textbf{0.89} \\ \hline
\end{tabular}%
}
\caption{Comparing the ranking performance of different recommender system models with D2Rec for Ciao.}
\label{table5}
\vspace{-8mm}
\end{table}
\subsubsection{Baselines}
Given that our framework focuses on mitigating confounding bias in an explicit feedback setting by leveraging auxiliary network information, we consider the following three types of baselines that represent the benchmark works in the respective categories:

\noindent{\textbf{Social Recommender Systems.}} D2Rec leverages auxiliary network information to account for confounders and better capture the user and item latent features. We, therefore, compare D2Rec with benchmarks for social recommendations.

\textit{Social Matrix Factorization - SocialMF}~\cite{jamali2010matrix} leverages users' social networks to model their preferences. This work considers social relations among users by adding propagation of each relation into the matrix factorization model.
%The work considers the social relations between users by adding the propagation of each relation into the matrix factorization model.

\textit{Graphrec: Graph Neural Networks for Social Recommendation - 
Graphrec}~\cite{fan2019graph} models social information with a Graph Neural Network. It leverages user-user social networks to model user latent features. It also uses the opinions associated with the user-item interactions network to model item latent features. 

\textit{ConsisRec: Enhancing GNN for Social Recommendation via Consistent Neighbor Aggregation}~\cite{yang2021consisrec} It is the state-of-the-art method in social recommendation. This model introduces the social inconsistency problem in social recommender systems and proposes to solve this problem by sampling-based attention mechanism.

\noindent{\textbf{Traditional Recommender Systems.}} The second set of baselines are traditional recommender systems.

\textit{Neural Collaborative Filtering - NeuMF}~\cite{he2017neural} is a state-of-the-art collaborative filtering model with neural network architecture. The authors model the user and item features through embedding layers and leverage a multi-layer perceptron to learn the user-item interactions for predicting ratings.

\textit{Probabilistic Matrix Factorization - PMF}~\cite{mnih2008probabilistic} models the user preference matrix as a product of two low-rank user and item matrices generated from Gaussian distributions. The approach also uses adaptive priors over the user and item latent features for model complexity.

\noindent{\textbf{Causal Recommender Systems.}} We compare D2Rec with the baselines that leverage causal inference for debiasing recommender systems with explicit feedback and DICE~\cite{zheng2021disentangling} which is a causal disentanglement framework proposed under an implicit feedback setting.

\textit{Inverse Propensity Score Matrix Factorization - IPS-MF}~\cite{liang2016causal} makes use of inverse propensity scores to alleviate the selection bias generated from the exposure data. The authors assign a probability to each possible user-item pair to ensure that each user-item interaction can be observed. Then, poisson factorization is used to compute propensity scores from the observed exposure matrix, and weighted matrix factorization is leveraged to correct for bias.

\textit{Causal Inference for Recommender Systems - CIRS}~\cite{wang2020causal} models the item exposure to a user as the treatment and the observed ratings as the outcomes. To measure the hidden confounders, this work uses a substitute by generating predicted exposure with the help of Poisson Factorization. It is used to compute the ratings by including the contribution of confounders for each user. This approach is integrated into a standard matrix factorization model.

\textit{Disentangling User Interest and Conformity for Recommendation with Causal Embedding}~\cite{zheng2021disentangling} considers the implicit recommender system setting. It considers the user interest and the conformity as the causes for the observed clicks and proposes a disentanglement framework for debiased recommendations. We only compare this baseline against the ranking prediction since the MAE and MSE metrics are not valid for implicit recommender systems.

% \usepackage{multirow}
% \usepackage{graphicx}
\begin{table}[t!]
\centering
%\small
\resizebox{0.65\textwidth}{!}{%
\begin{tabular}{|c|cccccccc|}
\hline
\multirow{3}{*}{\textbf{Model}} & \multicolumn{8}{c|}{\textbf{Popularity Debiased Test Sets}} \\ \cline{2-9} 
 & \multicolumn{2}{c|}{\textbf{Popularity=2}} & \multicolumn{2}{c|}{\textbf{Popularity=3}} & \multicolumn{2}{c|}{\textbf{Popularity=5}} & \multicolumn{2}{c|}{\textbf{Popularity=10}} \\ \cline{2-9} 
 & \multicolumn{1}{c|}{\textbf{MAE}} & \multicolumn{1}{c|}{\textbf{MSE}} & \multicolumn{1}{c|}{\textbf{MAE}} & \multicolumn{1}{c|}{\textbf{MSE}} & \multicolumn{1}{c|}{\textbf{MAE}} & \multicolumn{1}{c|}{\textbf{MSE}} & \multicolumn{1}{c|}{\textbf{MAE}} & \textbf{MSE} \\ \hline
\textbf{SocialMF~\cite{jamali2010matrix}} & \multicolumn{1}{c|}{1.36} & \multicolumn{1}{c|}{3.89} & \multicolumn{1}{c|}{1.05} & \multicolumn{1}{c|}{2.32} & \multicolumn{1}{c|}{0.91} & \multicolumn{1}{c|}{1.61} & \multicolumn{1}{c|}{0.87} & 1.47 \\ \hline
\textbf{GraphRec~\cite{fan2019graph}} & \multicolumn{1}{c|}{0.64} & \multicolumn{1}{c|}{0.88} & \multicolumn{1}{c|}{0.68} & \multicolumn{1}{c|}{0.90} & \multicolumn{1}{c|}{0.72} & \multicolumn{1}{c|}{0.97} & \multicolumn{1}{c|}{0.75} & 1.05 \\ \hline
\textbf{ConsisRec~\cite{yang2021consisrec}} & \multicolumn{1}{c|}{\textbf{0.63}} & \multicolumn{1}{c|}{\textbf{0.85}} & \multicolumn{1}{c|}{0.60} & \multicolumn{1}{c|}{0.84} & \multicolumn{1}{c|}{0.56} & \multicolumn{1}{c|}{0.78} & \multicolumn{1}{c|}{0.52} & 0.70 \\ \hline
\textbf{NeuMF~\cite{he2017neural}} & \multicolumn{1}{c|}{1.31} & \multicolumn{1}{c|}{2.75} & \multicolumn{1}{c|}{1.27} & \multicolumn{1}{c|}{2.61} & \multicolumn{1}{c|}{1.25} & \multicolumn{1}{c|}{2.51} & \multicolumn{1}{c|}{1.23} & 2.48 \\ \hline
\textbf{PMF~\cite{mnih2008probabilistic}} & \multicolumn{1}{c|}{1.25} & \multicolumn{1}{c|}{2.92} & \multicolumn{1}{c|}{1.03} & \multicolumn{1}{c|}{2.01} & \multicolumn{1}{c|}{0.93} & \multicolumn{1}{c|}{1.60} & \multicolumn{1}{c|}{0.92} & 1.56 \\ \hline
\textbf{IPS-MF~\cite{liang2016causal}} & \multicolumn{1}{c|}{1.01} & \multicolumn{1}{c|}{1.68} & \multicolumn{1}{c|}{0.98} & \multicolumn{1}{c|}{1.59} & \multicolumn{1}{c|}{0.97} & \multicolumn{1}{c|}{1.57} & \multicolumn{1}{c|}{0.95} & 1.53 \\ \hline
\textbf{CIRS~\cite{wang2020causal}} & \multicolumn{1}{c|}{1.08} & \multicolumn{1}{c|}{3.36} & \multicolumn{1}{c|}{0.74} & \multicolumn{1}{c|}{1.90} & \multicolumn{1}{c|}{0.49} & \multicolumn{1}{c|}{0.94} & \multicolumn{1}{c|}{0.35} & 0.55 \\ \hline
\textbf{D2Rec} & \multicolumn{1}{c|}{0.72} & \multicolumn{1}{c|}{1.49} & \multicolumn{1}{c|}{\textbf{0.49}} & \multicolumn{1}{c|}{\textbf{0.82}} & \multicolumn{1}{c|}{\textbf{0.34}} & \multicolumn{1}{c|}{\textbf{0.44}} & \multicolumn{1}{c|}{\textbf{0.22}} & \textbf{0.25} \\ \hline
\end{tabular}%
}
\caption{Comparing the prediction performance of different recommender system models with D2Rec for Epinions.}
\label{table2}
\vspace{-2mm}
\end{table}

% \usepackage{multirow}
% \usepackage{graphicx}
\begin{table}[t!]
\centering
%\small
\resizebox{0.65\textwidth}{!}{%
\begin{tabular}{|c|cccccccc|}
\hline
\multirow{3}{*}{\textbf{Model}} & \multicolumn{8}{c|}{\textbf{Popularity Debiased Test Sets}} \\ \cline{2-9} 
 & \multicolumn{2}{c|}{\textbf{Popularity=2}} & \multicolumn{2}{c|}{\textbf{Popularity=3}} & \multicolumn{2}{c|}{\textbf{Popularity=5}} & \multicolumn{2}{c|}{\textbf{Popularity=10}} \\ \cline{2-9} 
 & \multicolumn{1}{c|}{\textbf{MAE}} & \multicolumn{1}{c|}{\textbf{MSE}} & \multicolumn{1}{c|}{\textbf{MAE}} & \multicolumn{1}{c|}{\textbf{MSE}} & \multicolumn{1}{c|}{\textbf{MAE}} & \multicolumn{1}{c|}{\textbf{MSE}} & \multicolumn{1}{c|}{\textbf{MAE}} & \textbf{MSE} \\ \hline
\textbf{SocialMF~\cite{jamali2010matrix}} & \multicolumn{1}{c|}{1.40} & \multicolumn{1}{c|}{4.20} & \multicolumn{1}{c|}{1.00} & \multicolumn{1}{c|}{2.20} & \multicolumn{1}{c|}{0.78} & \multicolumn{1}{c|}{1.10} & \multicolumn{1}{c|}{0.72} & 0.95 \\ \hline
\textbf{GraphRec~\cite{fan2019graph}} & \multicolumn{1}{c|}{0.62} & \multicolumn{1}{c|}{0.72} & \multicolumn{1}{c|}{0.67} & \multicolumn{1}{c|}{0.80} & \multicolumn{1}{c|}{0.71} & \multicolumn{1}{c|}{0.84} & \multicolumn{1}{c|}{0.74} & 0.92 \\ \hline
\textbf{ConsisRec~\cite{yang2021consisrec}} & \multicolumn{1}{c|}{0.59} & \multicolumn{1}{c|}{\textbf{0.60}} & \multicolumn{1}{c|}{0.54} & \multicolumn{1}{c|}{0.58} & \multicolumn{1}{c|}{0.50} & \multicolumn{1}{c|}{0.46} & \multicolumn{1}{c|}{0.43} & 0.32 \\ \hline
\textbf{NeuMF~\cite{he2017neural}} & \multicolumn{1}{c|}{1.23} & \multicolumn{1}{c|}{2.43} & \multicolumn{1}{c|}{1.20} & \multicolumn{1}{c|}{2.31} & \multicolumn{1}{c|}{1.16} & \multicolumn{1}{c|}{2.15} & \multicolumn{1}{c|}{1.13} & 2.09 \\ \hline
\textbf{PMF~\cite{mnih2008probabilistic}} & \multicolumn{1}{c|}{1.21} & \multicolumn{1}{c|}{2.75} & \multicolumn{1}{c|}{0.94} & \multicolumn{1}{c|}{1.72} & \multicolumn{1}{c|}{0.80} & \multicolumn{1}{c|}{1.15} & \multicolumn{1}{c|}{0.78} & 1.08 \\ \hline
\textbf{IPS-MF~\cite{liang2016causal}} & \multicolumn{1}{c|}{1.12} & \multicolumn{1}{c|}{2.02} & \multicolumn{1}{c|}{1.09} & \multicolumn{1}{c|}{1.94} & \multicolumn{1}{c|}{1.05} & \multicolumn{1}{c|}{1.80} & \multicolumn{1}{c|}{1.03} & 1.73 \\ \hline
\textbf{CIRS~\cite{wang2020causal}} & \multicolumn{1}{c|}{0.92} & \multicolumn{1}{c|}{3.10} & \multicolumn{1}{c|}{0.52} & \multicolumn{1}{c|}{1.44} & \multicolumn{1}{c|}{0.28} & \multicolumn{1}{c|}{0.56} & \multicolumn{1}{c|}{0.17} & 0.24 \\ \hline
\textbf{D2Rec} & \multicolumn{1}{c|}{\textbf{0.40}} & \multicolumn{1}{c|}{0.72} & \multicolumn{1}{c|}{\textbf{0.21}} & \multicolumn{1}{c|}{\textbf{0.29}} & \multicolumn{1}{c|}{\textbf{0.10}} & \multicolumn{1}{c|}{\textbf{0.07}} & \multicolumn{1}{c|}{\textbf{0.04}} & \textbf{0.02} \\ \hline
\end{tabular}%
}
\caption{Comparing the prediction performance of different recommender system models with D2Rec for Ciao.}
\label{table3}
\vspace{-8mm}
\end{table}
\subsection{Performance Comparison (RQ.1)}
We compare the different baseline models with D2Rec on two real-world datasets, Epinions and Ciao. We split the test set into multiple test sets such that in each set, we ensure the number of ratings per item is the same for all items. Table~\ref{table2} (for Epinions) and Table~\ref{table3} (for Ciao) demonstrates the performance comparison across test sets with the number of ratings per item ranging from 2 through 10 for the different prediction performance metrics (MSE and MAE), and Table~\ref{table4} (for Epinions) and Table~\ref{table5} (for Ciao) shows the performance against different ranking performance metrics (Hit-Ratio@10 and NDCG@10). %The traditional recommender system baselines are represented in black; social recommendation baselines are shown in red, and causal perspective-based recommender system baselines are shown in blue. 
We have the following observations regarding \textbf{RQ.1}:
\begin{itemize}[leftmargin=*]
    \item Overall, D2Rec consistently yields the best performance among all datasets. For instance, it leads to more accurate rating prediction over the best baseline w.r.t. MSE/MAE. The results indicate that disentangling the factors with the help of social network information facilitates in debiasing recommendations. D2Rec also outperforms the other baselines in ranking performance, achieving the highest NDCG@K and Hit-Ratio@K scores across different test sets.
    \item Among the three types of baselines, the causal recommender systems serve as the strongest baselines in most cases, justifying that accounting for the underlying causal model is effective for debiasing recommender systems with explicit feedback. Causal recommender system models outperform traditional baselines in both rating prediction and ranking. D2Rec further outperforms the causal baselines because (1) it leverages social network information as a substitute for the unobserved confounders and (2) D2Rec disentangles the user and item representations into independent factors (exposure factors, rating factors, or confounders) as shown in the causal graph (see Fig.~\ref{causal_graph}). We also observe that DICE~\cite{zheng2021disentangling} outperforms other baselines showing that causal disentanglement is beneficial for producing debiased recommendations. However, D2Rec outperforms DICE due to two reasons. First, D2Rec uses network information to learn a better substitute for the hidden confounders, which DICE does not. Second, the performance discrepancy can result from different design purposes of DICE and D2Rec. DICE is designed specifically for the implicit feedback setting where the resulting causal graph is different from D2Rec's explicit feedback setting.
    \item It is observed that the performance of the traditional recommender baselines PMF and NeuMF, can drop significantly when items with extremely low popularity dominate the test sets (e.g., when the number of ratings per item is 2). This occurs because these models assume that ratings are missing at random, leading to learn biased embeddings. As a result, their prediction performance deteriorates on debiased test sets. Thus, these baselines are outperformed by D2Rec and the causal baselines in most of the cases.
   \item The error rate for the GraphRec baseline %starts at the lowest error rate but 
   increases as the number of ratings per item increases.
  We believe this occurs because of a problem prominent to Graph Neural Networks (GNNs) -- over-smoothing~\cite{cai2020note}.
   GNNs aggregate feature vectors from the neighbors of a node and combine them with the node's features to compute its representations.
   In addition, item popularity follows a long-tailed distribution~\cite{yang2018unbiased}.
  Over-smoothing can make GraphRec overfit the unpopular items as they dominate the population. Although popular items have more ratings in the training set, the computation of their representations considers their k-hop neighbors where there are enormous unpopular items. We follow~\cite{fan2019graph} to use three-layer GNNs in the GraphRec implementation. 
  This explains why GraphRec performs well when unpopular items dominate the test set while its performance drops when the number of ratings per item increases in the debiased test sets. Since ConsisRec utilizes a more restrictive sampling strategy based on consistency scores between users and utilizes an attention mechanism to select consistent relations, it overcomes the over-smoothing problem.
%   As a result, interacting nodes tend to have quite similar representations. Due to over-smoothing, nodes might have similar embeddings but not the same label (rating/exposure), which would cause mislabeling of them.
   %
   SocialMF does well as it does not use a GNN to learn embeddings but learns user features by only incorporating the information of 1-hop neighbors in both user-social and user-item networks.
\end{itemize}
\subsection{Ablation Study (RQ.2)}
\begin{table}[]
\centering
%\small
\resizebox{0.75\textwidth}{!}{%
\begin{tabular}{|c|cccccccc|}
\hline
\multirow{3}{*}{\textbf{Model}} & \multicolumn{8}{c|}{\textbf{Popularity Debiased Test Sets}} \\ \cline{2-9} 
 & \multicolumn{2}{c|}{\textbf{Popularity=2}} & \multicolumn{2}{c|}{\textbf{Popularity=3}} & \multicolumn{2}{c|}{\textbf{Popularity=5}} & \multicolumn{2}{c|}{\textbf{Popularity=10}} \\ \cline{2-9} 
 & \multicolumn{1}{c|}{\textbf{MAE}} & \multicolumn{1}{c|}{\textbf{MSE}} & \multicolumn{1}{c|}{\textbf{MAE}} & \multicolumn{1}{c|}{\textbf{MSE}} & \multicolumn{1}{c|}{\textbf{MAE}} & \multicolumn{1}{c|}{\textbf{MSE}} & \multicolumn{1}{c|}{\textbf{MAE}} & \textbf{MSE} \\ \hline
\textbf{D2Rec} & \multicolumn{1}{c|}{0.72} & \multicolumn{1}{c|}{1.49} & \multicolumn{1}{c|}{\textbf{0.49}} & \multicolumn{1}{c|}{\textbf{0.82}} & \multicolumn{1}{c|}{\textbf{0.34}} & \multicolumn{1}{c|}{\textbf{0.44}} & \multicolumn{1}{c|}{\textbf{0.22}} & \textbf{0.25} \\ \hline
\textbf{D2Rec w/o network embeddings} & \multicolumn{1}{c|}{\textbf{0.70}} & \multicolumn{1}{c|}{\textbf{1.30}} & \multicolumn{1}{c|}{0.53} & \multicolumn{1}{c|}{0.83} & \multicolumn{1}{c|}{0.41} & \multicolumn{1}{c|}{0.49} & \multicolumn{1}{c|}{0.30} & 0.34 \\ \hline
\textbf{D2Rec w/o disentanglement} & \multicolumn{1}{c|}{0.72} & \multicolumn{1}{c|}{1.67} & \multicolumn{1}{c|}{0.58} & \multicolumn{1}{c|}{1.03} & \multicolumn{1}{c|}{0.45} & \multicolumn{1}{c|}{0.50} & \multicolumn{1}{c|}{0.31} & 0.33 \\ \hline
\end{tabular}%
}
\caption{Analyzing (1) the effect of removing the disentanglement from D2Rec and (2) the effect of removing network embedding across different metrics for Epinions.}
\label{table6}
\vspace{-6mm}
\end{table}
% \usepackage{multirow}
% \usepackage{graphicx}
\begin{table}[]
\centering
%\small
\resizebox{0.75\textwidth}{!}{%
\begin{tabular}{|c|cccccccc|}
\hline
\multirow{3}{*}{\textbf{Model}} & \multicolumn{8}{c|}{\textbf{Popularity Debiased Test Sets}} \\ \cline{2-9} 
 & \multicolumn{2}{c|}{\textbf{Popularity=2}} & \multicolumn{2}{c|}{\textbf{Popularity=3}} & \multicolumn{2}{c|}{\textbf{Popularity=5}} & \multicolumn{2}{c|}{\textbf{Popularity=10}} \\ \cline{2-9} 
 & \multicolumn{1}{c|}{\textbf{MAE}} & \multicolumn{1}{c|}{\textbf{MSE}} & \multicolumn{1}{c|}{\textbf{MAE}} & \multicolumn{1}{c|}{\textbf{MSE}} & \multicolumn{1}{c|}{\textbf{MAE}} & \multicolumn{1}{c|}{\textbf{MSE}} & \multicolumn{1}{c|}{\textbf{MAE}} & \textbf{MSE} \\ \hline
\textbf{D2Rec} & \multicolumn{1}{c|}{0.40} & \multicolumn{1}{c|}{0.72} & \multicolumn{1}{c|}{\textbf{0.21}} & \multicolumn{1}{c|}{\textbf{0.29}} & \multicolumn{1}{c|}{\textbf{0.10}} & \multicolumn{1}{c|}{\textbf{0.07}} & \multicolumn{1}{c|}{\textbf{0.04}} & \textbf{0.02} \\ \hline
\textbf{D2Rec w/o network embeddings} & \multicolumn{1}{c|}{\textbf{0.40}} & \multicolumn{1}{c|}{\textbf{0.60}} & \multicolumn{1}{c|}{0.26} & \multicolumn{1}{c|}{0.32} & \multicolumn{1}{c|}{0.16} & \multicolumn{1}{c|}{0.12} & \multicolumn{1}{c|}{0.10} & 0.04 \\ \hline
\textbf{D2Rec w/o disentanglement} & \multicolumn{1}{c|}{0.43} & \multicolumn{1}{c|}{0.80} & \multicolumn{1}{c|}{0.30} & \multicolumn{1}{c|}{0.54} & \multicolumn{1}{c|}{0.20} & \multicolumn{1}{c|}{0.18} & \multicolumn{1}{c|}{0.14} & 0.10 \\ \hline
\end{tabular}%
}
\caption{Analyzing (1) the effect of removing the disentanglement from D2Rec and (2) the effect of removing network embedding across different metrics for Ciao.}
\label{table7}
\vspace{-8mm}
\end{table}

To assess the contribution of the different components of D2Rec concerning the performance, we conduct the following experiments. We consider two variants of D2Rec, one which is trained on non-network embeddings removing the effect induced by the network information namely, \textit{D2Rec w/o network embeddings}. The second variant of D2Rec is utilizes the network information but doesn't undergo any form of disentanglement. We denote this variant by \textit{D2Rec w/o disentanglement}. We conduct the same experiments across popularity debiased test sets where the item popularity ranges from 2-10. The results obtained can be seen in Table~\ref{table6} for Epinions and Table~\ref{table7} for Ciao. As observed, D2Rec performs the best when both the network information and the disentanglement are considered. Among the two variants, it is observed that D2Rec benefits more from the disentanglement module when compared to the network information. We believe such a pattern is observed because the network information aids in accounting for a specific type of confounders, i.e., the user conformity. In contrast, the disentanglement module acts as the base of learning specific components of the user and item representations that helps debias the recommender system by introducing the contribution of confounders in rating prediction.

% As mentioned earlier, the disentangled factor $\gamma$ accounts for the factors contributing to the confounders present in the user and item latent features. To verify the importance of $\gamma$, we experimented by entirely removing it from our proposed model (D2Rec) and compared the error with the original model. The results available in Fig.~\ref{ablation_D2Rec} suggest that when we do not account for the confounders, then the learned factors cannot segregate between user preferences and confounders. As a result, the error rate for D2Rec without the confounding factor $\gamma$ is higher.
\section{Conclusion and Future work}
\label{conc}
This work aims to leverage network information to debias social recommendations with the aid of causal disentanglement. Recommender systems suffer from a range of biases which hurt their performance. A common approach to generate debiased recommendations is to pose the recommender system problem from a causal perspective. However, this setting has not been adapted for the social recommender systems. We first formulate the social recommender systems problem from a causal perspective. In this setting, the exposure of an item to the user is analogous to the treatment, and the rating of an item is analogous to the outcome. A user’s preferences are similar to or influenced by their social relations. Thus, the social relations can aid in mitigating confounding bias as they act as noisy measures of hidden confounders. Aside from the user conformity, aspects of confounding such as item popularity present in the network information is also captured in our method with the aid of causal disentanglement which unravels the learned representations into three independent factors. Each factor contributes to either predicting the exposure (treatment) or the rating (outcome) or accounts for the hidden confounders. 
%To ensure the learned factors are independent of each other we utilize Integral Probabilistic Metrics to measure how dissimilar two representations are. Also, to mitigate the bias captured in the confounders we leverage context-aware importance sampling weights. Furthermore, we observe that the causal baselines outperform most of the baselines. This shows that, leveraging the underlying causal model is effective for debiasing recommender systems. Aside from the user conformity, aspects of confounding such as item popularity present in the network information is also captured in our method with the aid of \textit{causal disentanglement} which unravels the learned
Empirical evaluations on two real-world datasets corroborate the effectiveness of D2Rec. By leveraging network information and causal disentanglement, D2Rec consistently outperforms state-of-the-art algorithms with remarkable improvements. The analysis of the different components of D2Rec reveals that the causal disentanglement is more important than the network information in debiasing the recommendation performance. 

D2Rec considers a unified representation for the confounders. A meaningful direction for future work is extending D2Rec to include more fine-grained confounders, such as user conformity and item popularity factors. Another possible direction would be to explore how D2Rec would perform with distrust networks. Overall, we believe causal disentanglement of the user and item latent representations opens new doors for understanding user-item interactions of recommender systems.
%This paper focuses on leveraging network information to account for hidden confounders that influence a user's exposure to the item and the mechanism through which a user rates the items. Inspired by methods leveraging disentangling factors to measure substitutes for confounders in causal inference, we develop a principled framework, Causal Disentanglement for Debiased Recommendations (D2Rec), that disentangles user and item latent features into three independent factors. These factors are responsible for exposure (treatment), ratings (outcome), and confounding bias. We compare and evaluate D2Rec against various state-of-the-art baselines. Results highlight the importance of network information and the learned disentangled factor for confounders. %The learned disentangled factors are of high quality and interpretability and can be used to explore novel applications.

%In future work, we will explore how signed networks can enhance disentanglement. Another promising direction would be to extend the current framework by analyzing and leveraging different confounding factors like user demographics and integrating them with the current approach to better estimate the confounders.

% \bibliographystyle{ACM-Reference-Format}


\begin{thebibliography}{44}

%%% the Web address.

\ifx \showCODEN    \undefined \def \showCODEN     #1{\unskip}     \fi
\ifx \showDOI      \undefined \def \showDOI       #1{#1}\fi
\ifx \showISBNx    \undefined \def \showISBNx     #1{\unskip}     \fi
\ifx \showISBNxiii \undefined \def \showISBNxiii  #1{\unskip}     \fi
\ifx \showISSN     \undefined \def \showISSN      #1{\unskip}     \fi
\ifx \showLCCN     \undefined \def \showLCCN      #1{\unskip}     \fi
\ifx \shownote     \undefined \def \shownote      #1{#1}          \fi
\ifx \showarticletitle \undefined \def \showarticletitle #1{#1}   \fi
\ifx \showURL      \undefined \def \showURL       {\relax}        \fi
% invisible to TeX
\providecommand\bibfield[2]{#2}
\providecommand\bibinfo[2]{#2}
\providecommand\natexlab[1]{#1}
\providecommand\showeprint[2][]{arXiv:#2}

\bibitem[Abdollahpouri et~al\mbox{.}(2017)]%
        {abdollahpouri2017controlling}
\bibfield{author}{\bibinfo{person}{Himan Abdollahpouri}, \bibinfo{person}{Robin
  Burke}, {and} \bibinfo{person}{Bamshad Mobasher}.}
  \bibinfo{year}{2017}\natexlab{}.
\newblock \showarticletitle{Controlling popularity bias in learning-to-rank
  recommendation}. In \bibinfo{booktitle}{\emph{Proceedings of the eleventh ACM
  conference on recommender systems}}. \bibinfo{pages}{42--46}.
\newblock


\bibitem[Abdollahpouri et~al\mbox{.}(2021)]%
        {abdollahpouri2021user}
\bibfield{author}{\bibinfo{person}{Himan Abdollahpouri},
  \bibinfo{person}{Masoud Mansoury}, \bibinfo{person}{Robin Burke},
  \bibinfo{person}{Bamshad Mobasher}, {and} \bibinfo{person}{Edward
  Malthouse}.} \bibinfo{year}{2021}\natexlab{}.
\newblock \showarticletitle{User-centered evaluation of popularity bias in
  recommender systems}. In \bibinfo{booktitle}{\emph{Proceedings of the 29th
  ACM Conference on User Modeling, Adaptation and Personalization}}.
  \bibinfo{pages}{119--129}.
\newblock


\bibitem[Baddeley(2010)]%
        {baddeley2010herding}
\bibfield{author}{\bibinfo{person}{Michelle Baddeley}.}
  \bibinfo{year}{2010}\natexlab{}.
\newblock \showarticletitle{Herding, social influence and economic
  decision-making: socio-psychological and neuroscientific analyses}.
\newblock \bibinfo{journal}{\emph{Philosophical Transactions of the Royal
  Society B: Biological Sciences}} \bibinfo{volume}{365},
  \bibinfo{number}{1538} (\bibinfo{year}{2010}), \bibinfo{pages}{281--290}.
\newblock


\bibitem[Bonner and Vasile(2018)]%
        {bonner2018causal}
\bibfield{author}{\bibinfo{person}{Stephen Bonner} {and}
  \bibinfo{person}{Flavian Vasile}.} \bibinfo{year}{2018}\natexlab{}.
\newblock \showarticletitle{Causal embeddings for recommendation}. In
  \bibinfo{booktitle}{\emph{Proceedings of the 12th ACM conference on
  recommender systems}}. \bibinfo{pages}{104--112}.
\newblock


\bibitem[Borges and Stefanidis(2021)]%
        {borges2021mitigating}
\bibfield{author}{\bibinfo{person}{Rodrigo Borges} {and}
  \bibinfo{person}{Kostas Stefanidis}.} \bibinfo{year}{2021}\natexlab{}.
\newblock \showarticletitle{On mitigating popularity bias in recommendations
  via variational autoencoders}. In \bibinfo{booktitle}{\emph{Proceedings of
  the 36th Annual ACM Symposium on Applied Computing}}.
  \bibinfo{pages}{1383--1389}.
\newblock


\bibitem[B{\"u}hlmann(2020)]%
        {buhlmann2020invariance}
\bibfield{author}{\bibinfo{person}{Peter B{\"u}hlmann}.}
  \bibinfo{year}{2020}\natexlab{}.
\newblock \showarticletitle{Invariance, causality and robustness}.
\newblock \bibinfo{journal}{\emph{Statist. Sci.}} \bibinfo{volume}{35},
  \bibinfo{number}{3} (\bibinfo{year}{2020}), \bibinfo{pages}{404--426}.
\newblock


\bibitem[Cai and Wang(2020)]%
        {cai2020note}
\bibfield{author}{\bibinfo{person}{Chen Cai} {and} \bibinfo{person}{Yusu
  Wang}.} \bibinfo{year}{2020}\natexlab{}.
\newblock \showarticletitle{A note on over-smoothing for graph neural
  networks}.
\newblock \bibinfo{journal}{\emph{arXiv preprint arXiv:2006.13318}}
  (\bibinfo{year}{2020}).
\newblock


\bibitem[Chen et~al\mbox{.}(2020)]%
        {chen2020bias}
\bibfield{author}{\bibinfo{person}{Jiawei Chen}, \bibinfo{person}{Hande Dong},
  \bibinfo{person}{Xiang Wang}, \bibinfo{person}{Fuli Feng},
  \bibinfo{person}{Meng Wang}, {and} \bibinfo{person}{Xiangnan He}.}
  \bibinfo{year}{2020}\natexlab{}.
\newblock \showarticletitle{Bias and Debias in Recommender System: A Survey and
  Future Directions}.
\newblock \bibinfo{journal}{\emph{arXiv preprint arXiv:2010.03240}}
  (\bibinfo{year}{2020}).
\newblock


\bibitem[Fan et~al\mbox{.}(2019)]%
        {fan2019graph}
\bibfield{author}{\bibinfo{person}{Wenqi Fan}, \bibinfo{person}{Yao Ma},
  \bibinfo{person}{Qing Li}, \bibinfo{person}{Yuan He}, \bibinfo{person}{Eric
  Zhao}, \bibinfo{person}{Jiliang Tang}, {and} \bibinfo{person}{Dawei Yin}.}
  \bibinfo{year}{2019}\natexlab{}.
\newblock \showarticletitle{Graph neural networks for social recommendation}.
  In \bibinfo{booktitle}{\emph{The World Wide Web Conference}}.
  \bibinfo{pages}{417--426}.
\newblock


\bibitem[Gopalan et~al\mbox{.}(2015)]%
        {gopalan2015scalable}
\bibfield{author}{\bibinfo{person}{Prem Gopalan}, \bibinfo{person}{Jake~M
  Hofman}, {and} \bibinfo{person}{David~M Blei}.}
  \bibinfo{year}{2015}\natexlab{}.
\newblock \showarticletitle{Scalable Recommendation with Hierarchical Poisson
  Factorization.}. In \bibinfo{booktitle}{\emph{UAI}}.
  \bibinfo{pages}{326--335}.
\newblock


\bibitem[Gretton et~al\mbox{.}(2012)]%
        {gretton2012kernel}
\bibfield{author}{\bibinfo{person}{Arthur Gretton}, \bibinfo{person}{Karsten~M
  Borgwardt}, \bibinfo{person}{Malte~J Rasch}, \bibinfo{person}{Bernhard
  Sch{\"o}lkopf}, {and} \bibinfo{person}{Alexander Smola}.}
  \bibinfo{year}{2012}\natexlab{}.
\newblock \showarticletitle{A kernel two-sample test}.
\newblock \bibinfo{journal}{\emph{The Journal of Machine Learning Research}}
  \bibinfo{volume}{13}, \bibinfo{number}{1} (\bibinfo{year}{2012}),
  \bibinfo{pages}{723--773}.
\newblock


\bibitem[Grover and Leskovec(2016)]%
        {grover2016node2vec}
\bibfield{author}{\bibinfo{person}{Aditya Grover} {and} \bibinfo{person}{Jure
  Leskovec}.} \bibinfo{year}{2016}\natexlab{}.
\newblock \showarticletitle{node2vec: Scalable feature learning for networks}.
  In \bibinfo{booktitle}{\emph{Proceedings of the 22nd ACM SIGKDD international
  conference on Knowledge discovery and data mining}}.
  \bibinfo{pages}{855--864}.
\newblock


\bibitem[Hassanpour and Greiner(2019a)]%
        {hassanpour2019counterfactual}
\bibfield{author}{\bibinfo{person}{Negar Hassanpour} {and}
  \bibinfo{person}{Russell Greiner}.} \bibinfo{year}{2019}\natexlab{a}.
\newblock \showarticletitle{CounterFactual Regression with Importance Sampling
  Weights.}. In \bibinfo{booktitle}{\emph{IJCAI}}. \bibinfo{pages}{5880--5887}.
\newblock


\bibitem[Hassanpour and Greiner(2019b)]%
        {hassanpour2019learning}
\bibfield{author}{\bibinfo{person}{Negar Hassanpour} {and}
  \bibinfo{person}{Russell Greiner}.} \bibinfo{year}{2019}\natexlab{b}.
\newblock \showarticletitle{Learning disentangled representations for
  counterfactual regression}. In \bibinfo{booktitle}{\emph{International
  Conference on Learning Representations}}.
\newblock


\bibitem[He et~al\mbox{.}(2017)]%
        {he2017neural}
\bibfield{author}{\bibinfo{person}{Xiangnan He}, \bibinfo{person}{Lizi Liao},
  \bibinfo{person}{Hanwang Zhang}, \bibinfo{person}{Liqiang Nie},
  \bibinfo{person}{Xia Hu}, {and} \bibinfo{person}{Tat-Seng Chua}.}
  \bibinfo{year}{2017}\natexlab{}.
\newblock \showarticletitle{Neural collaborative filtering}. In
  \bibinfo{booktitle}{\emph{Proceedings of the 26th international conference on
  world wide web}}. \bibinfo{pages}{173--182}.
\newblock


\bibitem[Hu et~al\mbox{.}(2020)]%
        {hu2020graph}
\bibfield{author}{\bibinfo{person}{Linmei Hu}, \bibinfo{person}{Siyong Xu},
  \bibinfo{person}{Chen Li}, \bibinfo{person}{Cheng Yang},
  \bibinfo{person}{Chuan Shi}, \bibinfo{person}{Nan Duan},
  \bibinfo{person}{Xing Xie}, {and} \bibinfo{person}{Ming Zhou}.}
  \bibinfo{year}{2020}\natexlab{}.
\newblock \showarticletitle{Graph neural news recommendation with unsupervised
  preference disentanglement}. In \bibinfo{booktitle}{\emph{Proceedings of the
  58th Annual Meeting of the Association for Computational Linguistics}}.
  \bibinfo{pages}{4255--4264}.
\newblock


\bibitem[Jamali and Ester(2010)]%
        {jamali2010matrix}
\bibfield{author}{\bibinfo{person}{Mohsen Jamali} {and} \bibinfo{person}{Martin
  Ester}.} \bibinfo{year}{2010}\natexlab{}.
\newblock \showarticletitle{A matrix factorization technique with trust
  propagation for recommendation in social networks}. In
  \bibinfo{booktitle}{\emph{Proceedings of the fourth ACM conference on
  Recommender systems}}. \bibinfo{pages}{135--142}.
\newblock


\bibitem[Li et~al\mbox{.}(2021)]%
        {li2021causal}
\bibfield{author}{\bibinfo{person}{Qian Li}, \bibinfo{person}{Xiangmeng Wang},
  {and} \bibinfo{person}{Guandong Xu}.} \bibinfo{year}{2021}\natexlab{}.
\newblock \showarticletitle{Be Causal: De-biasing Social Network Confounding in
  Recommendation}.
\newblock \bibinfo{journal}{\emph{arXiv preprint arXiv:2105.07775}}
  (\bibinfo{year}{2021}).
\newblock


\bibitem[Liang et~al\mbox{.}(2016)]%
        {liang2016causal}
\bibfield{author}{\bibinfo{person}{Dawen Liang}, \bibinfo{person}{Laurent
  Charlin}, {and} \bibinfo{person}{David~M Blei}.}
  \bibinfo{year}{2016}\natexlab{}.
\newblock \showarticletitle{Causal inference for recommendation}. In
  \bibinfo{booktitle}{\emph{Causation: Foundation to Application, Workshop at
  UAI. AUAI}}.
\newblock


\bibitem[Liu et~al\mbox{.}(2020)]%
        {liu2020explainable}
\bibfield{author}{\bibinfo{person}{Ninghao Liu}, \bibinfo{person}{Yong Ge},
  \bibinfo{person}{Li Li}, \bibinfo{person}{Xia Hu}, \bibinfo{person}{Rui
  Chen}, {and} \bibinfo{person}{Soo-Hyun Choi}.}
  \bibinfo{year}{2020}\natexlab{}.
\newblock \showarticletitle{Explainable recommender systems via resolving
  learning representations}. In \bibinfo{booktitle}{\emph{Proceedings of the
  29th ACM International Conference on Information \& Knowledge Management}}.
  \bibinfo{pages}{895--904}.
\newblock


\bibitem[Ma et~al\mbox{.}(2009)]%
        {ma2009trust}
\bibfield{author}{\bibinfo{person}{Nan Ma}, \bibinfo{person}{Ee-Peng Lim},
  \bibinfo{person}{Viet-An Nguyen}, \bibinfo{person}{Aixin Sun}, {and}
  \bibinfo{person}{Haifeng Liu}.} \bibinfo{year}{2009}\natexlab{}.
\newblock \showarticletitle{Trust relationship prediction using online product
  review data}. In \bibinfo{booktitle}{\emph{Proceedings of the 1st ACM
  international workshop on Complex networks meet information \& knowledge
  management}}. \bibinfo{pages}{47--54}.
\newblock


\bibitem[Marlin and Zemel(2009)]%
        {marlin2009collaborative}
\bibfield{author}{\bibinfo{person}{Benjamin~M Marlin} {and}
  \bibinfo{person}{Richard~S Zemel}.} \bibinfo{year}{2009}\natexlab{}.
\newblock \showarticletitle{Collaborative prediction and ranking with
  non-random missing data}. In \bibinfo{booktitle}{\emph{Proceedings of the
  third ACM conference on Recommender systems}}. \bibinfo{pages}{5--12}.
\newblock


\bibitem[Marsden and Friedkin(1993)]%
        {marsden1993network}
\bibfield{author}{\bibinfo{person}{Peter~V Marsden} {and}
  \bibinfo{person}{Noah~E Friedkin}.} \bibinfo{year}{1993}\natexlab{}.
\newblock \showarticletitle{Network studies of social influence}.
\newblock \bibinfo{journal}{\emph{Sociological Methods \& Research}}
  \bibinfo{volume}{22}, \bibinfo{number}{1} (\bibinfo{year}{1993}),
  \bibinfo{pages}{127--151}.
\newblock


\bibitem[Mnih and Salakhutdinov(2008)]%
        {mnih2008probabilistic}
\bibfield{author}{\bibinfo{person}{Andriy Mnih} {and} \bibinfo{person}{Russ~R
  Salakhutdinov}.} \bibinfo{year}{2008}\natexlab{}.
\newblock \showarticletitle{Probabilistic matrix factorization}. In
  \bibinfo{booktitle}{\emph{Advances in neural information processing
  systems}}. \bibinfo{pages}{1257--1264}.
\newblock


\bibitem[M{\"u}ller(1997)]%
        {muller1997integral}
\bibfield{author}{\bibinfo{person}{Alfred M{\"u}ller}.}
  \bibinfo{year}{1997}\natexlab{}.
\newblock \showarticletitle{Integral probability metrics and their generating
  classes of functions}.
\newblock \bibinfo{journal}{\emph{Advances in Applied Probability}}
  \bibinfo{volume}{29}, \bibinfo{number}{2} (\bibinfo{year}{1997}),
  \bibinfo{pages}{429--443}.
\newblock


\bibitem[Nema et~al\mbox{.}(2020)]%
        {nema2020untangle}
\bibfield{author}{\bibinfo{person}{Preksha Nema}, \bibinfo{person}{Alexandros
  Karatzoglou}, {and} \bibinfo{person}{Filip Radlinski}.}
  \bibinfo{year}{2020}\natexlab{}.
\newblock \showarticletitle{Untangle: Critiquing Disentangled Recommendations}.
\newblock  (\bibinfo{year}{2020}).
\newblock


\bibitem[Peters et~al\mbox{.}(2017)]%
        {peters2017elements}
\bibfield{author}{\bibinfo{person}{Jonas Peters}, \bibinfo{person}{Dominik
  Janzing}, {and} \bibinfo{person}{Bernhard Sch{\"o}lkopf}.}
  \bibinfo{year}{2017}\natexlab{}.
\newblock \bibinfo{booktitle}{\emph{Elements of causal inference: foundations
  and learning algorithms}}.
\newblock \bibinfo{publisher}{The MIT Press}.
\newblock


\bibitem[Qian et~al\mbox{.}(2021)]%
        {qian2021intent}
\bibfield{author}{\bibinfo{person}{Tieyun Qian}, \bibinfo{person}{Yile Liang},
  \bibinfo{person}{Qing Li}, \bibinfo{person}{Xuan Ma}, \bibinfo{person}{Ke
  Sun}, {and} \bibinfo{person}{Zhiyong Peng}.} \bibinfo{year}{2021}\natexlab{}.
\newblock \showarticletitle{Intent Disentanglement and Feature Self-supervision
  for Novel Recommendation}.
\newblock \bibinfo{journal}{\emph{arXiv preprint arXiv:2106.14388}}
  (\bibinfo{year}{2021}).
\newblock


\bibitem[Rosenbaum and Rubin(1983)]%
        {rosenbaum1983central}
\bibfield{author}{\bibinfo{person}{Paul~R Rosenbaum} {and}
  \bibinfo{person}{Donald~B Rubin}.} \bibinfo{year}{1983}\natexlab{}.
\newblock \showarticletitle{The central role of the propensity score in
  observational studies for causal effects}.
\newblock \bibinfo{journal}{\emph{Biometrika}} \bibinfo{volume}{70},
  \bibinfo{number}{1} (\bibinfo{year}{1983}), \bibinfo{pages}{41--55}.
\newblock


\bibitem[Schnabel et~al\mbox{.}(2016)]%
        {schnabel2016recommendations}
\bibfield{author}{\bibinfo{person}{Tobias Schnabel}, \bibinfo{person}{Adith
  Swaminathan}, \bibinfo{person}{Ashudeep Singh}, \bibinfo{person}{Navin
  Chandak}, {and} \bibinfo{person}{Thorsten Joachims}.}
  \bibinfo{year}{2016}\natexlab{}.
\newblock \showarticletitle{Recommendations as treatments: Debiasing learning
  and evaluation}. In \bibinfo{booktitle}{\emph{international conference on
  machine learning}}. PMLR, \bibinfo{pages}{1670--1679}.
\newblock


\bibitem[Schulz(1998)]%
        {schulz1998randomized}
\bibfield{author}{\bibinfo{person}{Kenneth~F Schulz}.}
  \bibinfo{year}{1998}\natexlab{}.
\newblock \showarticletitle{Randomized controlled trials}.
\newblock \bibinfo{journal}{\emph{Clinical obstetrics and gynecology}}
  \bibinfo{volume}{41}, \bibinfo{number}{2} (\bibinfo{year}{1998}),
  \bibinfo{pages}{245--256}.
\newblock


\bibitem[Shadish et~al\mbox{.}(2008)]%
        {shadish2008can}
\bibfield{author}{\bibinfo{person}{William~R Shadish},
  \bibinfo{person}{Margaret~H Clark}, {and} \bibinfo{person}{Peter~M Steiner}.}
  \bibinfo{year}{2008}\natexlab{}.
\newblock \showarticletitle{Can nonrandomized experiments yield accurate
  answers? A randomized experiment comparing random and nonrandom assignments}.
\newblock \bibinfo{journal}{\emph{Journal of the American statistical
  association}} \bibinfo{volume}{103}, \bibinfo{number}{484}
  (\bibinfo{year}{2008}), \bibinfo{pages}{1334--1344}.
\newblock


\bibitem[Sinha et~al\mbox{.}(2001)]%
        {sinha2001comparing}
\bibfield{author}{\bibinfo{person}{Rashmi~R Sinha}, \bibinfo{person}{Kirsten
  Swearingen}, {et~al\mbox{.}}} \bibinfo{year}{2001}\natexlab{}.
\newblock \showarticletitle{Comparing recommendations made by online systems
  and friends.}
\newblock \bibinfo{journal}{\emph{DELOS}}  \bibinfo{volume}{106}
  (\bibinfo{year}{2001}).
\newblock


\bibitem[Tang et~al\mbox{.}(2012)]%
        {tang2012mtrust}
\bibfield{author}{\bibinfo{person}{Jiliang Tang}, \bibinfo{person}{Huiji Gao},
  {and} \bibinfo{person}{Huan Liu}.} \bibinfo{year}{2012}\natexlab{}.
\newblock \showarticletitle{mTrust: Discerning multi-faceted trust in a
  connected world}. In \bibinfo{booktitle}{\emph{Proceedings of the fifth ACM
  international conference on Web search and data mining}}.
  \bibinfo{pages}{93--102}.
\newblock


\bibitem[Wang et~al\mbox{.}(2020)]%
        {wang2020causal}
\bibfield{author}{\bibinfo{person}{Yixin Wang}, \bibinfo{person}{Dawen Liang},
  \bibinfo{person}{Laurent Charlin}, {and} \bibinfo{person}{David~M Blei}.}
  \bibinfo{year}{2020}\natexlab{}.
\newblock \showarticletitle{Causal Inference for Recommender Systems}. In
  \bibinfo{booktitle}{\emph{Fourteenth ACM Conference on Recommender Systems}}.
  \bibinfo{pages}{426--431}.
\newblock


\bibitem[Wei et~al\mbox{.}(2021)]%
        {wei2021model}
\bibfield{author}{\bibinfo{person}{Tianxin Wei}, \bibinfo{person}{Fuli Feng},
  \bibinfo{person}{Jiawei Chen}, \bibinfo{person}{Ziwei Wu},
  \bibinfo{person}{Jinfeng Yi}, {and} \bibinfo{person}{Xiangnan He}.}
  \bibinfo{year}{2021}\natexlab{}.
\newblock \showarticletitle{Model-agnostic counterfactual reasoning for
  eliminating popularity bias in recommender system}. In
  \bibinfo{booktitle}{\emph{Proceedings of the 27th ACM SIGKDD Conference on
  Knowledge Discovery \& Data Mining}}. \bibinfo{pages}{1791--1800}.
\newblock


\bibitem[Yang et~al\mbox{.}(2018)]%
        {yang2018unbiased}
\bibfield{author}{\bibinfo{person}{Longqi Yang}, \bibinfo{person}{Yin Cui},
  \bibinfo{person}{Yuan Xuan}, \bibinfo{person}{Chenyang Wang},
  \bibinfo{person}{Serge Belongie}, {and} \bibinfo{person}{Deborah Estrin}.}
  \bibinfo{year}{2018}\natexlab{}.
\newblock \showarticletitle{Unbiased offline recommender evaluation for
  missing-not-at-random implicit feedback}. In
  \bibinfo{booktitle}{\emph{Proceedings of the 12th ACM Conference on
  Recommender Systems}}. \bibinfo{pages}{279--287}.
\newblock


\bibitem[Yang et~al\mbox{.}(2021b)]%
        {yang2021consisrec}
\bibfield{author}{\bibinfo{person}{Liangwei Yang}, \bibinfo{person}{Zhiwei
  Liu}, \bibinfo{person}{Yingtong Dou}, \bibinfo{person}{Jing Ma}, {and}
  \bibinfo{person}{Philip~S Yu}.} \bibinfo{year}{2021}\natexlab{b}.
\newblock \showarticletitle{Consisrec: Enhancing gnn for social recommendation
  via consistent neighbor aggregation}. In
  \bibinfo{booktitle}{\emph{Proceedings of the 44th international ACM SIGIR
  conference on Research and development in information retrieval}}.
  \bibinfo{pages}{2141--2145}.
\newblock


\bibitem[Yang et~al\mbox{.}(2021a)]%
        {yang2021top}
\bibfield{author}{\bibinfo{person}{Mengyue Yang}, \bibinfo{person}{Quanyu Dai},
  \bibinfo{person}{Zhenhua Dong}, \bibinfo{person}{Xu Chen},
  \bibinfo{person}{Xiuqiang He}, {and} \bibinfo{person}{Jun Wang}.}
  \bibinfo{year}{2021}\natexlab{a}.
\newblock \showarticletitle{Top-N Recommendation with Counterfactual User
  Preference Simulation}. In \bibinfo{booktitle}{\emph{Proceedings of the 30th
  ACM International Conference on Information \& Knowledge Management}}.
  \bibinfo{pages}{2342--2351}.
\newblock


\bibitem[Yu et~al\mbox{.}(2017)]%
        {yu2017selection}
\bibfield{author}{\bibinfo{person}{Hsiang-Fu Yu}, \bibinfo{person}{Mikhail
  Bilenko}, {and} \bibinfo{person}{Chih-Jen Lin}.}
  \bibinfo{year}{2017}\natexlab{}.
\newblock \showarticletitle{Selection of negative samples for one-class matrix
  factorization}. In \bibinfo{booktitle}{\emph{Proceedings of the 2017 SIAM
  International Conference on Data Mining}}. SIAM, \bibinfo{pages}{363--371}.
\newblock


\bibitem[Zafarani et~al\mbox{.}(2014)]%
        {zafarani2014social}
\bibfield{author}{\bibinfo{person}{Reza Zafarani},
  \bibinfo{person}{Mohammad~Ali Abbasi}, {and} \bibinfo{person}{Huan Liu}.}
  \bibinfo{year}{2014}\natexlab{}.
\newblock \bibinfo{booktitle}{\emph{Social media mining: an introduction}}.
\newblock \bibinfo{publisher}{Cambridge University Press}.
\newblock


\bibitem[Zhang et~al\mbox{.}(2021)]%
        {zhang2021causal}
\bibfield{author}{\bibinfo{person}{Yang Zhang}, \bibinfo{person}{Fuli Feng},
  \bibinfo{person}{Xiangnan He}, \bibinfo{person}{Tianxin Wei},
  \bibinfo{person}{Chonggang Song}, \bibinfo{person}{Guohui Ling}, {and}
  \bibinfo{person}{Yongdong Zhang}.} \bibinfo{year}{2021}\natexlab{}.
\newblock \showarticletitle{Causal Intervention for Leveraging Popularity Bias
  in Recommendation}.
\newblock \bibinfo{journal}{\emph{arXiv preprint arXiv:2105.06067}}
  (\bibinfo{year}{2021}).
\newblock


\bibitem[Zheng et~al\mbox{.}(2021)]%
        {zheng2021disentangling}
\bibfield{author}{\bibinfo{person}{Yu Zheng}, \bibinfo{person}{Chen Gao},
  \bibinfo{person}{Xiang Li}, \bibinfo{person}{Xiangnan He},
  \bibinfo{person}{Yong Li}, {and} \bibinfo{person}{Depeng Jin}.}
  \bibinfo{year}{2021}\natexlab{}.
\newblock \showarticletitle{Disentangling User Interest and Conformity for
  Recommendation with Causal Embedding}. In
  \bibinfo{booktitle}{\emph{Proceedings of the Web Conference 2021}}.
  \bibinfo{pages}{2980--2991}.
\newblock


\bibitem[Zhu et~al\mbox{.}(2021)]%
        {zhu2021popularity}
\bibfield{author}{\bibinfo{person}{Ziwei Zhu}, \bibinfo{person}{Yun He},
  \bibinfo{person}{Xing Zhao}, \bibinfo{person}{Yin Zhang},
  \bibinfo{person}{Jianling Wang}, {and} \bibinfo{person}{James Caverlee}.}
  \bibinfo{year}{2021}\natexlab{}.
\newblock \showarticletitle{Popularity-opportunity bias in collaborative
  filtering}. In \bibinfo{booktitle}{\emph{Proceedings of the 14th ACM
  International Conference on Web Search and Data Mining}}.
  \bibinfo{pages}{85--93}.
\newblock


\end{thebibliography}

% \appendix

\end{document}